\documentclass[traditabstract]{aa} 

\usepackage{natbib}
\bibpunct{(}{)}{;}{a}{}{,}

\usepackage{graphicx}
\usepackage{subcaption}
\usepackage{txfonts}
\newcommand{\halpha}{H$\alpha$}

\newcommand{\icoone}{$I_{\rm CO(1-0)}$}
\newcommand{\icotwo}{$I_{\rm CO(2-1)}$}
\newcommand{\lsun}{$L_\odot$}
\newcommand{\msun}{$M_\odot$}
\newcommand{\mi}{$\mu$m}
\newcommand{\kms}{km~s$^{-1}$}

\newcommand{\mmol} {\hbox{$M_{{\rm H}_2}$}}

\newcommand{\mdust} {\hbox{$M_{{\rm dust}}$}}
\newcommand{\mgas} {\hbox{$M_{{\rm gas}}$}}
\newcommand{\mstar} {\hbox{$M_{{\rm \star}}$}}

\newcommand{\mhi}   {\hbox{$M_{\rm HI}$}}

\newcommand{\sfrhalpha}{SFR$_{\rm H\alpha}$}
\newcommand{\sfrfuv}{SFR$_{\rm FUV}$}

\newcommand{\sigmamol}{$\Sigma_{\rm mol}$}

\newcommand{\sigmamstar}{$\Sigma_{{\rm M}_\star}$}
\newcommand{\sigmasfr}{$\Sigma_{\rm SFR}$}
\newcommand{\sigmasfrfuv}{$\Sigma_{{\rm SFR}_{\rm FUV}}$}
\newcommand{\sigmasfrhalpha}{$\Sigma_{{\rm SFR}_{\rm H\alpha}}$}

\newcommand{\shi}{S$_{\rm HI}$}
\newcommand{\spitzer}{{\it Spitzer} }

\newcommand{\taudep}{$\tau_{\rm dep}$}
\newcommand{\taudepfuv}{$\tau_{\rm dep,\ FUV}$}
\newcommand{\taudephalpha}{$\tau_{\rm dep,\ H\alpha}$}

\begin{document}

   \title{Star formation and gas in the minor merger UGC\,10214}


   \author{D. Rosado-Belza\inst{1,2,3}
          \and
          U. Lisenfeld\inst{1,4} 
          \and
          J. Hibbard\inst{5}
          \and
          K. Kniermann\inst{6,7}\thanks{NSF Astronomy and Astrophysics Postdoctoral Fellow}
          \and
          J. Ott\inst{8}
          \and
          S. Verley\inst{1,4}
          \and
          M. Boquien\inst{9}
           \and
          T. Jarrett\inst{10}
          \and
          C. K. Xu \inst{11,12}
          }
          
   \institute{Departamento de Fí­sica Teórica y del Cosmos, Facultad de Ciencias, University of Granada,
              Av. Fuentenueva s/n, 18071 Granada\\
              \email{darobel@correo.ugr.es}
         \and
             Instituto de Astrof\'{i}sica de Canarias,
             V\'{i}a L\'{a}ctea s/n, 38205 La Laguna, Spain
         \and
             Departamento de Astrof\'{i}sica, University of La Laguna,
             38206 La Laguna, Spain
         \and
             Instituto Carlos I, Facultad de Ciencias, University of Granada,
              Av. Fuentenueva s/n, 18071 Granada, Spain
         \and
         National Radio Astronomy Observatory, Charlottesville, VA 22903, USA
         \and
          School of Earth \& Space Exploration, Arizona State University, 550 E. Tyler Mall, Room PSF-686 (P.O. Box 876004), TEMPE, AZ 85287-6004
          \and
          NSF Astronomy and Astrophysics Postdoctoral Fellow
          \and
          National Radio Astronomy Observatory, 1003 Lopezville Road, Socorro, NM 87801, USA
          \and
          Unidad de Astronomía, Fac. Cs. Básicas, Universidad de Antofagasta, Avda. U. de Antofagasta 02800, Antofagasta, Chile
          \and
          Astrophysics, Cosmology and Gravity Centre, Astronomy Department, University of Cape Town, Private Bag X3, Rondebosch 7701, South Africa 
          \and
              National Astronomical Observatories, Chinese Academy of Sciences, Beijing 1000012, China
    \and
    South American centre for Astronomy, CAS, Camino El Observatorio 1515, Las Condes, Santiago, Chile
             }

   \date{}

 
    \abstract
  {
%
  Minor mergers play a crucial role in galaxy evolution. UGC\,10214 (the ``Tadpole Galaxy'') is a prime example of this process in which a dwarf galaxy has interacted with a large spiral galaxy $\sim$250\,Myr ago and produced a perturbed disk and a giant tidal tail. We  use a multiwavelength dataset, partly from new observations (\halpha, HI, and CO) and partly from archival data, in order to study the present and past star formation rate (SFR) and its relation to the  gas and stellar mass at a spatial resolution down to 4\,kpc. UGC\,10214 is a very massive (stellar mass \mstar = $1.28\times 10^{11}$\,\msun) galaxy with a low gas fraction (\mgas/\mstar = 0.24), a high molecular gas fraction (\mmol/\mhi = 0.4) and a modest SFR (2 -- 5~\msun\,yr$^{-1}$). The global SFR compared to its stellar mass places UGC\,10214 on the galaxy main sequence (MS). The comparison of the molecular gas mass and current  SFR gives a molecular gas depletion time of about $\sim$ 2\,Gyr (based on \halpha), comparable to those of normal spiral galaxies. Both from a comparison of the \halpha\ emission, tracing the current SFR, and far-ultraviolet (FUV) emission, tracing the recent SFR during the past tens of Myr, as well as from spectral energy distribution (SED) fitting with CIGALE, we find that the SFR has increased by a factor of about 2 -- 3 during the recent past. This increase is particularly noticeable in the centre of the galaxy where a pronounced peak of the \halpha\ emission is visible. A pixel-to-pixel comparison of the SFR, molecular gas mass and stellar mass shows that the central region has had a depressed FUV-traced SFR, both compared to the molecular gas and the stellar mass, whereas the \halpha-traced SFR shows a normal level. The atomic and molecular gas distribution is asymmetric, but the position-velocity diagram along the major axis shows a pattern of regular rotation. We conclude that the minor merger has most likely caused variations in the SFR in the past resulting in a moderate increase of the SFR, but it has not perturbed the gas significantly so that the molecular depletion time remains normal.
  }

   \maketitle
%

\section{Introduction}

The question of how galaxies, in particular their star formation (SF) activity, evolve along cosmic time and how they build up their stellar mass, \mstar, has been the centre of numerous studies in the recent literature. Data from large galaxy surveys have established a relation between the star formation rate (SFR) and stellar mass of galaxies \citep[e.g.][]{brinchmann04} which holds out to high redshift, albeit with a different normalisation \citep[e.g.][]{noeske07}. This  relation shows a bimodality between active star-forming galaxies (``blue cloud'') and quiescent galaxies (``red sequence'') with a smaller galaxy density in the intermediate, possibly transitioning region (``green valley''). The relation between the SFR and \mstar\ of the star-forming galaxies is also referred to as the main sequence (MS)  of galaxies and can be taken as the ``norm'' whereas galaxies above and below have an excess or deficiency of SF for their stellar mass. It is still an open question how galaxies evolve along and across the MS and what processes are mainly responsible for their transition towards the red sequence. Various processes have been suggested, including gravitational interaction (see below), the presence of an active galactic nucleus \citep[AGN, e.g.][]{ellison16, argudo16} and gas consumption or perturbation \citep[e.g.][]{saintonge16, 2017A&A...607A.110L}.

Gravitational interaction is an important process for the evolution of galaxies in clusters \citep{1980ApJ...236..351D, 1996Natur.379..613M}, groups \citep{1992ApJ...399..353H, 2017A&A...607A.110L}, triplets \citep{2015MNRAS.447.1399D, argudo15} and pairs \citep{2010MNRAS.407.1514E, argudo15}. Galaxies with nearby companions have shown alterations in their morphologies, star formation rates (SFRs) and gas reservoirs. In the most extreme cases, two or more galaxies may collide and their evolution may be severely altered through galaxy mergers. Major mergers between two similar mass galaxies may result in the formation of a single, massive elliptical galaxy \citep[][]{toomre78, schweizer82, 2006ApJ...652..270B}. Although less dramatic, minor mergers between unequal mass galaxies \citep[mass ratio $\lesssim0.3$,][]{bournaud05} might also greatly affect the evolution of both galaxies, in some cases leading to the formation of tidal tails and tidal dwarf galaxies \citep{2006A&A...456..481B, 2012MNRAS.419...70K}. The way in which two galaxies collide will determine the particular outcome of the remnant. To reach an insight of the consequences of a peculiar collision, the study of specific minor mergers, as an example of a simple collision scene, is necessary.

The Tadpole Galaxy (UGC\,10214, VV\,29, or Arp\,188) is a massive galaxy which has experienced a minor merger (where the remnant of the intruder galaxy represents less than 10\%  of the Tadpole's gas mass, see Sect. 4) several hundreds of Myr ago. The encounter has distorted the Tadpole's disk and caused gas to be ejected in a long ($\sim$110\,kpc) tidal tail. \citet{briggs01} studied the atomic hydrogen (HI)  gas distribution and kinematics of the  system and  identified three main  dynamical components: VV\,29a, which corresponds to the Tadpole's disk, VV\,29b, the tidal tail, and VV\,29c, the dwarf galaxy. \citet{tran03} and \citet{degrijs03} analysed images taken with the ACS/HST  and found a large number of young star cluster in the tail and the disk. The ages of these clusters, which range from less than 10\,Myr to 150 -- 200\,Myr, are younger than their estimate of the dynamical age of the interaction of $\sim$250\,Myr, based on the length of the tail and expected ejection velocity of the gas (close to rotation velocity of 400\,\kms). \citet{jarrett06} examined the star formation (SF) in UGC\,10214 using optical and infrared imaging. They reveal several infrared-bright hot spots in the spiral arms and little SF in the
central region. They measured a global SFR of 2 to 4\,\msun\,yr$^{-1}$.

These previous studies have shown that the Tadpole Galaxy is an excellent object to investigate the effect of a minor interaction on the interstellar medium (ISM) and the SF in the host galaxy. However, so far, data for  the molecular gas as a crucial ingredient for SF, as well as multiwavelength data allowing an analysis of the star formation history in this object have been missing. In order to fill this gap,  we carry out a multiwavelength analysis of UGC\,10214, including a large set of archival images from ultraviolet (UV) to submillimeter wavelengths, new HI data at a higher spatial resolution than previous data \citep{briggs01}, new CO data tracing the molecular gas, and an \halpha\ image that shows the distribution of the current SF. The main goal of this paper is to better understand the present and past SF and its relation to the neutral (atomic and molecular) gas in this galaxy that has suffered a minor interaction $\sim$250\,Myr ago in order to evaluate the impact of this event on the evolution of the galaxy.

This paper is organised as follows. In Section 2, we present the complete set of data used in our study. Section 3 gives a  description of the three methods employed in our study: aperture photometry of the disk of the galaxy as a whole and divided in three different regions; a spectral energy distribution (SED) modelling using Code Investigating GALaxy Emission (CIGALE), and a pixel-by-pixel analysis of the three regions defined for the aperture photometry. Section 4 presents the results of the distribution and kinematics of the atomic and molecular gas and of the SFR and the stellar mass. Additionally, we study the relation between these quantities, both on a global and on a spatially resolved scale. In Section 5 we discuss the results of the previous section, trying to reconstruct the  star formation history of UGC\,10214 and discuss its future evolution. Finally, we present a brief summary and our conclusions in Section 6.
We adopt a distance of $134.2\pm 9.4$\,Mpc \citep{mould00} for this galaxy.


\section{Data}

An ample set of data is available to study UGC\,10214, covering a broad wavelength range from the ultraviolet (UV) to the radio. Part of the data (observations of \halpha, CO, and HI) are from our own observations and the rest are from public archives. In this section we describe the data and, when adequate, their reduction. In Table~\ref{tab:data} we present a summary of the available data and in Fig.~{\ref{fig:insets} we show a selection of the images used.

\begin{figure*}
\centering
\includegraphics[width=\linewidth]{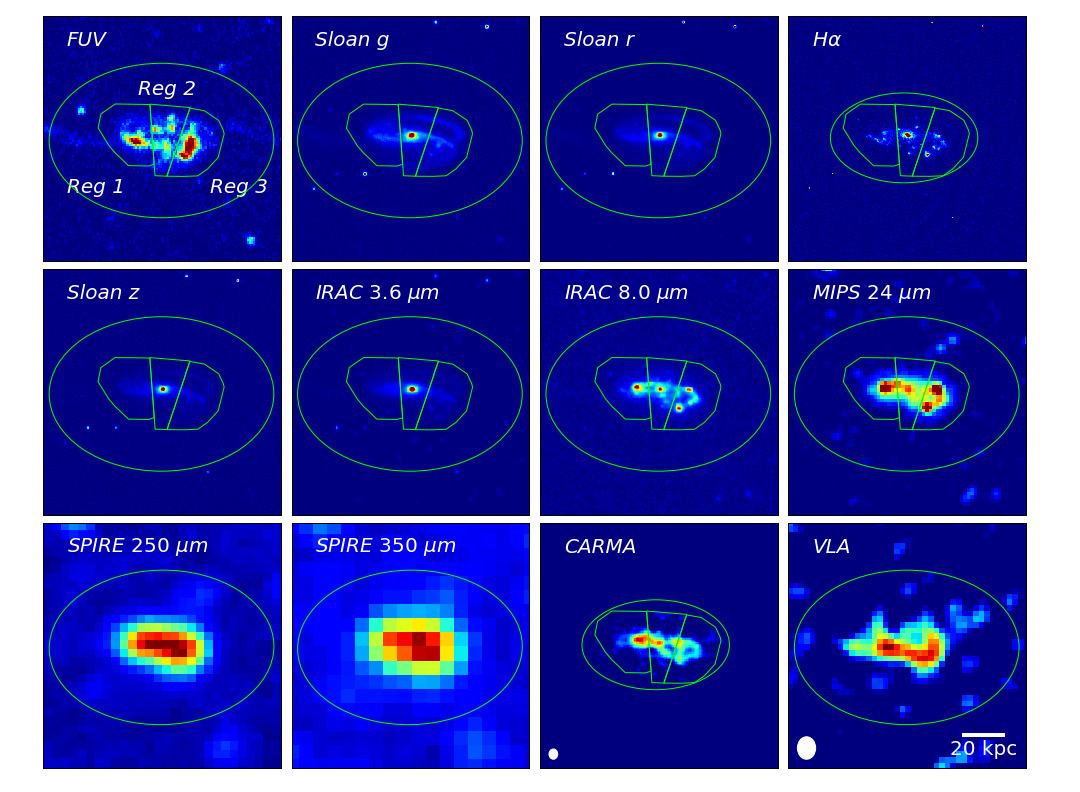}
\caption{A representative selection of the images used in our study, together with the  apertures used for the three regions (the polygonal shapes) and the disk (the elliptical shape). The elliptical apertures are selected in order to ensure the full coverage of the emission of the whole disk. The apertures for the regions in the disk cover three different star-forming regions observed in \halpha\ and UV images. The field of view of each image is 170\arcsec$\times$176\arcsec. The CARMA and VLA maps shows the velocity integrated intensities. The white ellipse (lower left corner) in the CARMA and VLA images represents the beam size for each image.}
\label{fig:insets}
\end{figure*}

\subsection{Ultraviolet}

To study the continuum UV emission from the young stars of the Tadpole Galaxy we use data from the \textit{GALaxy Evolution eXplorer} mission \textit{GALEX}. \textit{GALEX} is a space telescope providing images in two different ultraviolet bands: FUV, covering the wavelength range from 1350\,$\AA$ to 1750\,$\AA$, and NUV, covering the range from 1750\,$\AA$ to 2750\,$\AA$ \citep{martin05}.\

UGC\,10214 was observed in the Deep Imaging Survey (DIS) in both bands. We  retrieve the calibrated maps from the archive and subtract the background using the corresponding maps from the archive. We adopt a calibration error of $10\%$ for both images \citep{morrissey07}.\

\subsection{Optical}
\subsubsection{\textit{Sloan}}

We retrieve  images of UGC~10214 from \textit{Sloan Digital Sky Survey} (SDSS) \citep{york00} in the \textit{u'}, \textit{g'}, \textit{r'}, \textit{i'}, and \textit{z'} bands from the SDSS $9^{\rm th}$ Data Release (July $2012$). We adopt a calibration error of $1\%$ for the \textit{g'}, \textit{r'}, \textit{i'}, and \textit{z'} bands, and $2\%$ for the \textit{u'} band \citep{padmanabhan08}.

\subsubsection{\halpha}

The study of the ionised gas is performed using the narrow band \halpha\ image from Knierman et al. (in prep.). Observations to obtain \halpha\ images were taken with the Loral 2K CCD imager at the Lennon 1.8m Vatican Advanced Technology Telescope (VATT) on Mount Graham, Arizona on May 22, 2004. This imager has a $6\farcm{4}$ field of view with $0\farcs{42}$ per pixel. Narrow-band \halpha\ images were obtained at the VATT with an 88\,mm  Andover 3-cavity interference filter centered at 6780\,\AA. Integration time for each of the six \halpha\ images is 1200\,s.  To subtract continuum emission, this field was also observed with a Kron-Cousins R filter using integration times of 3$\times$300\,s. Images were reduced using standard IRAF\footnote{
IRAF is distributed by the National Optical Astronomy Observatory, which is operated by the Association of Universities for Research in Astronomy, Inc., under cooperative agreement with the National Science Foundation.} tasks.

To create images with only the emission lines, a scaled R band image was subtracted from the narrowband image after alignment using foreground stars. To determine an initial scaling factor, the ratio of integration time for individual frames (1200s/300s = 4) is multiplied by the ratio of filter widths (81$\rm \AA$/1186.35$\rm \AA$ = 0.068). This initial scaling factor then equals 0.272. We next determine the scaling factor by performing photometry of non-saturated stars in the \halpha\ and R band images.  By plotting the instrumental flux of both filters against each other, we determine a linear fit of f(\halpha$_{\rm inst}$) = 0.256  $\times$ f(R$_{\rm inst}$) + 185.79 with a correlation coefficient of $r=0.9965$. The slope of this fit gives the appropriate scaling factor between the R and \halpha\ images  which we adopt.

For calibration of the \halpha\ flux, spectrophotometric standard stars from the \citet{oke90} catalog were observed. Aperture photometry of these standards was compared to their absolute magnitudes. Absolute magnitudes for each spectrophotometric standard star were calculated by integrating their spectral energy distribution over the filter response function. We used a standard atmospheric extinction coefficient of 0.08 mag airmass$^{-1}$. Zeropoints were calculated by comparing the absolute magnitude in each filter with the instrumental magnitude from aperture photometry. For each night, the zeropoints from all standards (in this case, 4) were averaged. These zeropoints had a standard deviation of 0.016 mag.

Finally, we removed the contamination from the \textsc{[NII]} doublet at  6548,6583 $\rm \AA$ in the \halpha\ filter,
adopting a ratio of \textsc{[NII]/\halpha\ }= 0.094 \citep[from][]{tran03} .

\begin{table*}
	\caption[]{Summary of the images and data used in this study.}
\centering
\begin{tabular}{lccccc}
\hline
\hline
Telescope & Instrument/ & $\lambda_0$ ($\mu$m) & FWHM & $\triangle_{cal}$ & rms noise \\
     &  filter/line  & or  $\nu_0$ (GHz) &  & ($\%$) & of images\\
\hline
GALEX & FUV & $0.154$~\mi\ & $3\farcs{82}$ & $10$ & $2.4\times 10^{-5}$\,mJy\\
GALEX & NUV & $0.232$~\mi\ & $5\farcs{40}$ & $10$ & $5\times 10^{-5}$\,mJy\\ \hline
\textit{Sloan} & u' & $0.359$~\mi\ & $1\farcs{68}$ & $2$ & $3\times 10^{-4}$\,mJy\\
\textit{Sloan} & g' & $0.481$~\mi\ & $1\farcs{70}$ & $1$ & $1.1\times 10^{-3}$\,mJy\\
\textit{Sloan} & r' & $0.623$~\mi\ & $1\farcs{31}$ & $1$ & $1.6\times 10^{-3}$\,mJy\\
\textit{Sloan} & i' & $0.764$~\mi\ & $1\farcs{24}$ & $1$ & $2.3\times 10^{-3}$\,mJy\\
\textit{Sloan} & z' & $0.906$~\mi\ & $1\farcs{32}$ & $1$ & $2.5\times 10^{-3}$\,mJy\\ \hline
\textit{VATT} & \halpha\ & $0.677$~\mi\ & $0\farcs{80}$ & $3$ & $5\times 10^{-17}$\,erg cm$^{-2}$s$^{-1}$\\ \hline
\textit{Spitzer} & IRAC  & $3.58$~\mi\ & $2\farcs{16}$ & $10$ & $1.8\times 10^{-3}$\,mJy\\
\textit{Spitzer} & IRAC  & $4.52$~\mi\ & $2\farcs{14}$ & $10$ & $1.2\times 10^{-3}$\,mJy\\
\textit{Spitzer} & IRAC  & $5.72$~\mi\ & $1\farcs{83}$ & $10$ & $2.4\times 10^{-3}$\,mJy\\
\textit{Spitzer} & IRAC  & $7.90$~\mi\ & $2\farcs{13}$ & $10$ & $1.4\times 10^{-3}$\,mJy\\ \hline
\textit{Spitzer} & MIPS & $23.7$~\mi\ & $6\farcs{01}$ & $10$ & $6\times 10^{-3}$\,mJy\\ \hline
\textit{Herschel} & SPIRE  & $250$~\mi\ & $19\farcs{05}$ & $15$ & $0.6$\,mJy\\
\textit{Herschel} & SPIRE  & $350$~\mi\ & $27\farcs{18}$ & $15$ & $0.7$\,mJy\\
\textit{Herschel} & SPIRE  & $500$~\mi\ & $45\farcs{21}$ & $15$ & $0.7$\,mJy\\ \hline
CARMA & CO(1-0) &111.8~GHz & $6\farcs{2}\times7\farcs{3}$ & $10$ & $2$\,mJy\,beam$^{-1}$\\ \hline
IRAM 30m & CO(1-0) & 111.8~GHz &  $22\arcsec$ & $15$ & --\\ 
IRAM 30m     &  CO(2-1)  & 223.6~GHz & $11\arcsec$ & $25$ & --\\\hline
VLA & HI & 1.45~GHz & $13\arcsec- 16\arcsec$ & $3$ & $0.09-0.10$\,mJy\,beam$^{-1}$\\
\hline
\label{tab:data}
\end{tabular}
\end{table*}

\subsection{Infrared}

\subsubsection{\textit{Spitzer}}

To carry out the study of the galaxy in the near and mid-infrared we use a set of images obtained by the \textit{Spitzer Space Telescope}, consisting of images at 3.6\,\mi, 4.5\,\mi, 5.8\,\mi, and 8.0\,\mi\ taken by IRAC  \citep[Infrared Array Camera,][]{fazio04}, as well as an image at 24\,\mi\ from MIPS  \citep[Multiband Imaging Photometer,][]{rieke04} (project ID: 185, IP C. Londsdale).  No other images from MIPS were available. The images were reduced with  the Super-Moscaic Pipeline and no further data reduction was necessary. The 3.6\,\mi\ and 4.5\,\mi\ images mainly provide information about the emission of the oldest star population of the galaxy, the  5.8\,\mi\ and 8.0\,\mi\ images about the emission from the polycyclic aromatic hydrocarbons (PAH) and the 24\,\mi\ image about the emission from the hot dust. For all these images we adopt a calibration error of $10\%$ (\citealt{fazio04} for IRAC and \citealt{rieke04} for MIPS).

\subsubsection{\textit{Herschel}}

To further study the dust emission of the Tadpole Galaxy we use images obtained by the Spectral and Photometric Imaging Receiver  \citep[SPIRE,][]{griffin10} aboard the \textit{Herschel Space Telescope}. Images at all three SPIRE bands (250\,\mi, 350\,\mi, and 500\,\mi) are available in the \textit{Herschel} archive. The data were observed in SpirePacsParallel mode (project ID: KPGT\_soliver\_1). The PACS images are not useful because UGC~10124 is too close to the image edge. We retrieve the SPIRE level-3 products which are mosaics obtained by merging all or a subset of contiguous observations. No further data reduction was necessary. Following the  instrument handbook\footnote{http://herschel.esac.esa.int/Docs/SPIRE/spire\_handbook.pdf}, we adopt a calibration error of $15\%$.

\subsection{Millimetre data}

We carry out two complementary sets of CO observations. The disk of UGC\,10214 was observed with the interferometer CARMA (Combined Array for Research in Millimeter-wave Astronomy), at a spatial resolution of $\sim$7\arcsec\ and a velocity resolution of 81\,\kms. In addition, a fully sampled map was obtained with the 30\,m telescope at Pico Veleta of the Institut de Radioastronomie Milimétrique (IRAM). This map has a coarse spatial resolution (22\arcsec), but a narrower velocity resolution ($\sim$2.5\,\kms). The velocities of both data sets are adapted to the definition of the VLA data (optical convention for the transformation of frequency to velocity and heliocentric reference system).

\subsubsection{CARMA}

We observed the Tadpole galaxy in the $^{12}$CO(1-0) 115\,GHz transition with the Combined Array for Research in Millimeter-wave Astronomy (CARMA) in a 9-pointing mosaic that covered its head and tail structure (Project code c0269). CARMA was a heterogeneous array that consisted of 6 antennas with 10.4\,m diameter and 9 antennas with 6.1\,m diameter. We obtained data in the D and E-antenna configurations, covering baselines between 8 and 150\,m. Twelve D-configuration tracks were taken between 2008 July 20 and 2008 August 29 and nine E-configuration tracks between 2008 September 18 and 2008 October 5, with on-source time variations between 1 and 5\,h. 

The correlation was set up to cover 468.75\,MHz spread over 15 channels, 31.25\,MHz wide, using Doppler corrections for the  Tadpole with a redshift of $z=0.031358$ relative to the rest frequency of the CO line at 115.27120\,GHz. This corresponds to 81.27\,km\,s$^{-1}$ wide channels over a total bandwidth of 1219\,km\,s$^{-1}$.

All data reduction was performed with the MIRIAD package\footnote{http://www.atnf.csiro.au/computing/software/miriad/}
\citep{sault95}. The data were edited and calibrated using sources 0927+390 or 3C273 for bandpass (15\,min on source), 1638+573  for complex gain/phase (with cycles of 20\,min on the Tadpole galaxy and 2.5\,min on the calibrator), and Mars (5\,min) for flux calibration respectively. All data were combined to produce final, mosaicked image cubes using natural weighting and joint deconvolution with a synthesized beam of $7\farcs{28}\times6\farcs{23}; PA= 67.1^{\circ}$. The rms noise in the field of view that is covered by all antenna primary beams of the heterogeneous array amounts to 2\,mJy. We adopt a calibration error of $10\%$.

We calculate the mass of the molecular gas from the CARMA CO fluxes as:

\begin{equation}
\left(\frac{M_{\rm H_2}}{M_{\odot}}\right) = 1.05 \cdot 10^4 \cdot \left(\frac{S}{\rm Jy\cdot km s^{-1}}\right) \cdot \left(\frac{D}{\rm Mpc}\right)^2\quad,
\label{eq:mh2_carma}
\end{equation}

 from \cite{bolatto13}, for the redshift of Tadpole, z = 0.0313. This expression is based on a conversion factor of $X_{\rm CO}=2\times10^{20}$\,cm$^{-2}$/(K\,km\,s$^{-2}$) and includes a helium fraction of 1.36.

\subsubsection{IRAM 30\,m}

We observed the CO(1--0) and CO(2--1) lines at their redshifted frequencies (111.8\,GHz and 223.6\,GHz, respectively) between August and October 2005 with the IRAM 30\,m  telescope on Pico Veleta at 8 position, offset by 9\arcsec\ and covering the entire disk of UGC\,10214. The map is fully sampled for CO(1--0) but undersampled for CO(2--1). Dual polarisation receivers were used at both frequencies with the 512$\times$1\,MHz filterbanks on the CO(1--0) line and the 256$\times$4\,MHz filterbanks on the CO(2--1). The observations were done in wobbler switching mode with a wobbler throw of 200\arcsec \ in azimuthal direction. Pointing was monitored on nearby quasars  every 60--90  minutes. During the observation period, the weather conditions were  generally good (with pointing better than 4\arcsec). The average system temperatures were $\sim$200\,K at 115\,GHz and $\sim$420\,K at 230\,GHz on the $T_{\rm A}^\star$ scale. At 111.8\,GHz (223.6\,GHz), the IRAM forward efficiency, $F_{\rm eff}$, was 0.95 (0.90), the  beam efficiency, $B_{\rm eff}$, was 0.75 (0.54), and the half-power beam size is 22$^{\prime\prime}$ (11$^{\prime\prime}$). All CO spectra are presented on the main beam temperature scale ($T_{\rm mb}$) which is defined as $T_{\rm mb} = (F_{\rm eff}/B_{\rm eff})\times T_{\rm A}^\star$.

The data were reduced in the standard way via the CLASS software in the GILDAS package\footnote{http://www.iram.fr/IRAMFR/GILDAS}. We first discarded poor scans and then subtracted a constant or linear baseline. We then averaged the spectra over the individual positions. 

For each spectrum, we visually determine the zero-level line widths, if detected. The velocity integrated spectra is calculated by summing the individual channels in between these limits. All positions in CO(1--0) are detected. For CO(2--1), two positions are not detected and we obtained  the  upper limit as

\begin{equation}
I_{\rm CO} < 3 \times \rm{rms} \times \sqrt{\delta \rm{V} \ \Delta V}\quad,
\end{equation}

\noindent where $\delta \rm{V}$ is the channel width and $\Delta$V the total line width for which we adopt the same value as for the detected CO(1--0) line at this position. In addition to the statistical error of the velocity integrated line intensities, a typical calibration error of  15\% for CO(1-0) and 25\% for CO(2-1) has to be taken into account. 

In Fig.~\ref{fig:iram_co_spectra}, we show the positions observed by the IRAM 30\,m telescope, overlaid over the CO map of CARMA, together with the CO(1--0) and CO(2--1) spectra. In Table~\ref{tab:ico_mmol}  the velocity integrated intensities for the various positions are listed.

\begin{figure}
  \centering
    \includegraphics[width=0.5\textwidth,trim=0.cm 0.cm 0.cm 0.cm,clip]{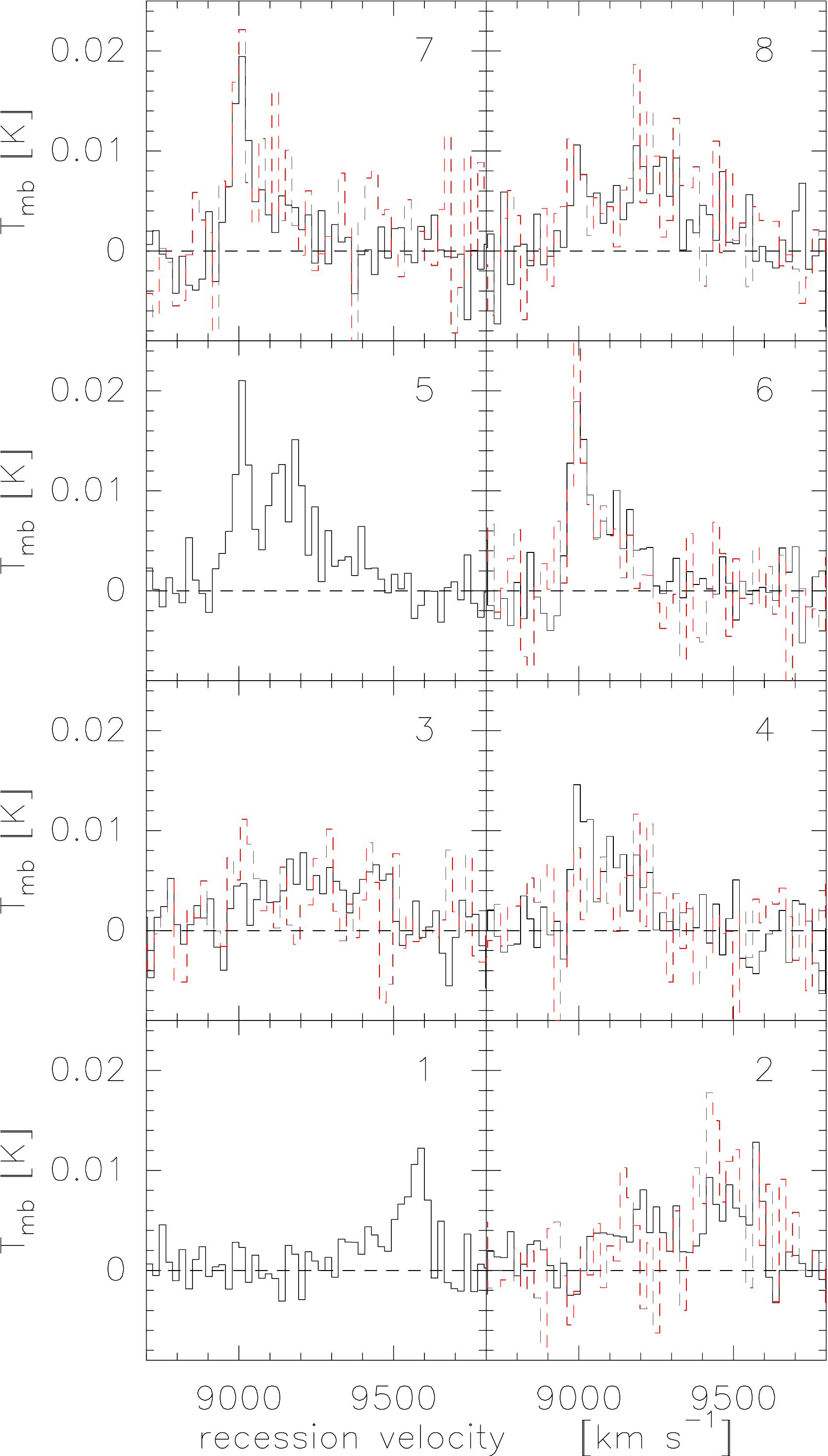}
   \includegraphics[width=0.4\textwidth,trim=0.cm 0cm 0.cm 0cm,clip]{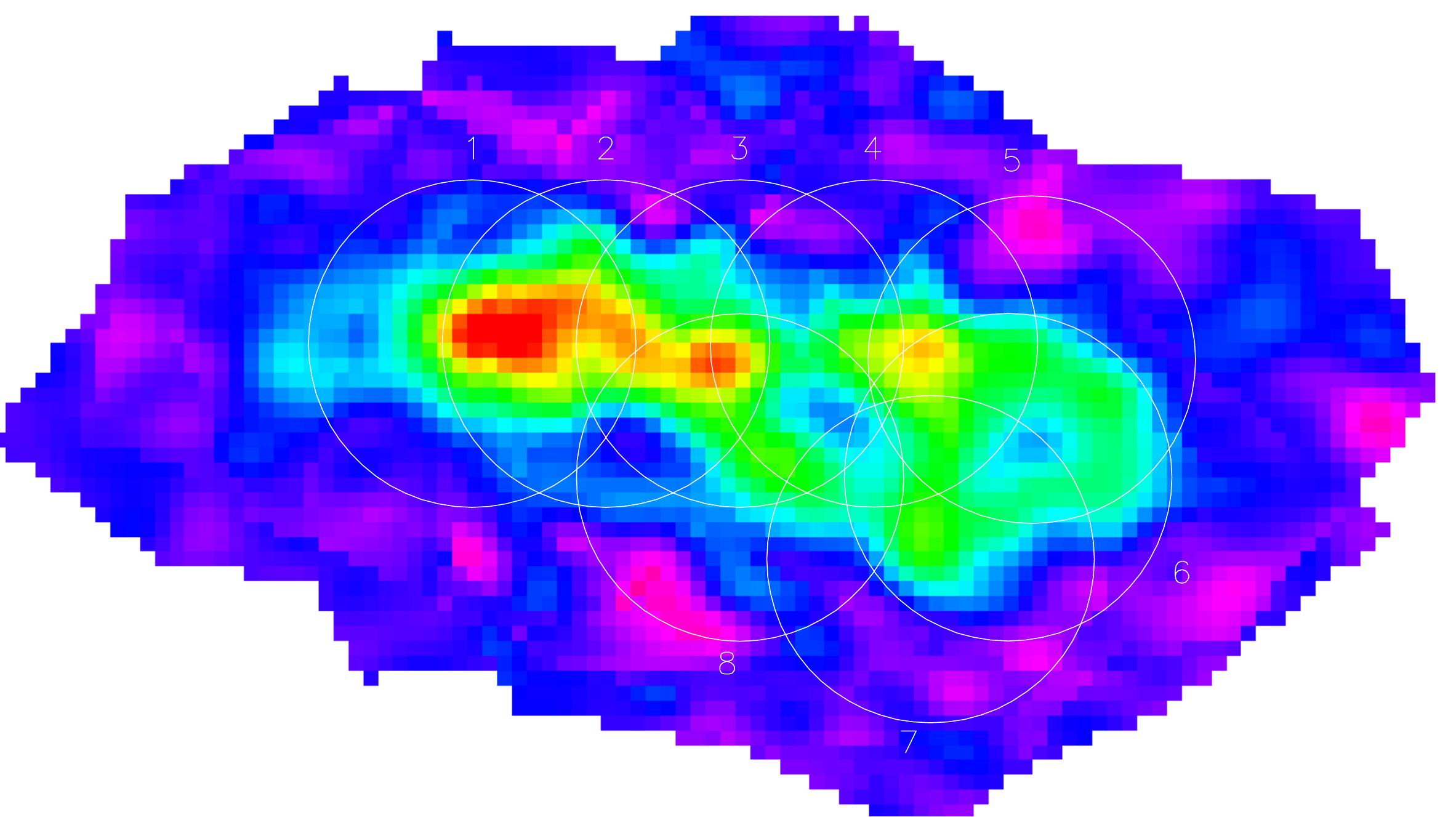}
  \caption{Upper panel: CO(1-0) (black, full line) and CO(2-1) (red, dashed line) spectra observed with the IRAM 30m telescope. 
  Only detected spectra
  are shown. Lower panel: The observed positions are 
  indicated on a CARMA CO(1-0) map. }
  \label{fig:iram_co_spectra}
\end{figure}

\begin{table}
\caption{CO(1-0) and CO(2-1) velocity integrated intensities from IRAM, and comparison to CARMA data.}           
\label{tab:ico_mmol}      
\centering          
\begin{tabular}{lccc}
\hline
\hline
Position & I$_{\rm CO10}$\tablefootmark{a} & I$_{\rm CO21}$\tablefootmark{a}  & $\frac{M_{\rm H_2,\ CARMA}}{M_{\rm H_2,\ IRAM}}$\tablefootmark{b}\\
  & (K \kms)  & (K \kms)   & \\
 \hline 
       1  & $1.5\pm0.2$ &  $<$ 1.1 &       $1.3\pm0.3$\\
     2  & $2.9\pm0.2$ &   $2.9\pm0.5$    &  $0.8\pm0.2$  \\
   3 & $2.4\pm0.3$ &  $1.9\pm0.4$   &  $0.9\pm0.2$\\
   4  & $2.0\pm0.2$ & $1.4\pm0.3$     &  $1.0\pm0.2$\\
       5 &  $3.7\pm0.3$ &  $<2.1$ &    $0.4\pm0.1$\\
  6 &  $2.4\pm0.2$ &  $2.1\pm0.3$ &    $0.7\pm0.2$\\
       7  & $2.0\pm0.2$ & $2.4\pm0.5$ &   $0.7\pm0.1$\\ 
8  &  $2.7\pm0.3$ &  $3.00\pm0.45$    &  $0.7\pm0.1$\\
\hline
\end{tabular}
\tablefoot{
\tablefoottext{a}{Velocity integrated CO(1-0) and CO(2-1) intensities observed at the positions labelled in 
Fig.~\ref{fig:iram_co_spectra} (in T$_{\rm mB}$).}
\tablefoottext{b}{Flux ratio between CARMA and IRAM at the individual position, derived as explained in Sect.~\ref{sec:comparison_iram_carma}.}
}
\end{table}

We calculate the molecular gas mass for the IRAM data from the CO(1-0) velocity integrated emission using the same value for $X_{\rm CO}=N(H_2)/I_{\rm CO(1-0)} = 2 \times 10^{20}$ cm$^{-2}$/(K \kms) as for the CARMA data and also including a helium fraction of 1.36.

\begin{eqnarray}
M_{\rm H_2} &=& 1.36  \cdot I_{\rm CO(1-0)}\frac{N(H_2)}{I_{\rm CO(1-0)}} D^2 \cdot \Omega \cdot 2 m_H \quad , \\
\left(\frac{M_{\rm H_2}}{M_{\odot}}\right) &=& 102\cdot 10^5 
\left(\frac{\Omega}{\rm arcsec^{2}}\right)  \left(\frac{I_{\rm CO(1-0)}}{\rm K\cdot km s^{-1}}\right)  \left(\frac{D}{\rm Mpc}\right)^2 \quad .
\label{eq:mh2_iram}
\end{eqnarray}

Here,  $\Omega$ is the area covered by the observations (i.e. for a single pointing  with a Gaussian beam $\Omega$ = 1.13 $\theta^{2}$ where $\theta$ is the half-power beam width, HPBW), $D$ is the distance and $2m_H$ is the mass of an $H_2$ molecule.}

\subsection{Radio (HI Data)}
\label{sec:hi_data}

The VLA observations were conducted on 2004 March 1 with 25 available antennas in the C-array configuration (maximum baseline of 3.4 km). At 21\,cm this provides angular resolutions (FWHM of the synthesised beam) of $\sim$13\arcsec\ -- 16\arcsec\ (depending on weighting, see below), which represents a factor of 2 -- 3 linear improvement over the 36\arcsec\ $\times$ 26\arcsec\ resolution of the Westerbork Radio Synthesis Telescope (WSRT) HI observations of \citet{briggs01}. The correlator setup was chosen considering the HI observations of \citet{briggs01}, which showed the entire HI emission in UGC\,10214 to span a total of 900\,\kms\ (velocities\footnote{Throughout this paper we use heliocentric velocities and the optical convention to transform frequency shifts to velocity.} from 8950 -- 9850\,\kms). At that time, the VLA correlator modes could either cover a wide velocity range at very low spectral resolution, or a narrower velocity range at higher spectral resolution. As a compromise we elected to use the 4IF mode, which allowed us to use different bandwidth codes for each IF pair. Both IF pairs were tuned to a central frequency corresponding to the peak tail emission (9420\,\kms). The first IF pair used bandwidth code 5 to provide 63 channels across a 1.56~MHz bandwidth (velocity coverage $\sim$9250 -- 9590\,\kms\  with a channel spacing of 5.5\,\kms). These data will be referred to as the \textit{narrowband data}. The second IF pair used bandwidth code 3 to provide 15 channels across a 6.25\,MHz bandwidth (velocity coverage $\sim$8806 -- 10034~\kms\ with a channel spacing of 88\,\kms). These data will be referred to as the \textit{wideband data}. The narrowband data provides higher velocity resolution to resolve the kinematics of the tidal gas (measured velocity dispersion of 17\,\kms\ from \citealt{briggs01}), but will not measure the full extent of the disk gas or intruding dwarf galaxy (VV\,29c), while the wideband data provides sufficient bandwidth to measure all HI associated with the system with line-free channels on either end, but at a very coarse velocity resolution.In the present paper we only use the wideband data of the disk of UGC\,10214. The wide- and narrowband data of the tidal tail will be presented and analysed in Knierman et al. (in prep.).

The flux calibrator 3C286 (1331+305) was observed during 15~min at the start of the observation, followed by pairs of observations on the phase calibrator (1634+627, 10~min per visit) and UGC\,10214 (45~min per visit). This phase calibrator was bright enough at 21\,cm (5\,Jy) to be also used as the bandpass calibrator. The total observation spanned 9 hours, with 6.5 hours on-source.

The data were reduced using the Astronomical Image Processing System (AIPS). Continuum subtraction was done based on the line-free channels using the AIPS task UVLIN. This is a visibility-domain technique that removes a linear fit to the line-free channels from each visibility. The data were imaged with various ``robust'' weighting factors \citep{briggs95} yielding FWHM beam sizes between $\sim$13\arcsec\ -- 16\arcsec\ and rms noises between 0.09 and 0.10\,mJy\,beam$^{-1}$. The resulting images are 3-dimensional cubes of HI intensity (in Jy\,beam$^{-1}$) as a function of R.A. and Dec; and velocity. In the present work we use the data cube with a ``robust'' weighting factor of +1, which provides  the coarsest spatial resolution (15\farcs{5}$\times$16\farcs{0}) but reflects most reliably the total flux and has a higher signal-to-noise ratio. We inspected the robust -1 map with a higher resolution (12\farcs{6}$\times$12\farcs{9})  for the presence of new features not visible in the robust +1 map but did not find any.

\begin{figure}
  \centering
  \includegraphics[width=0.5\textwidth,trim=0.cm 0.cm 0.cm 0.cm,clip]{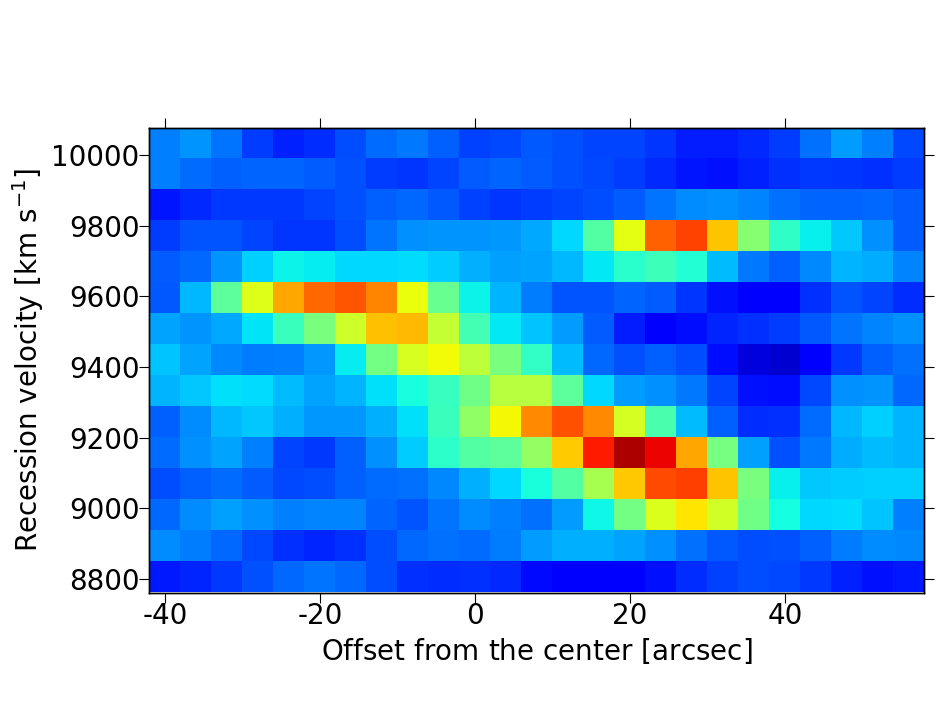}
  \caption{Position-velocity diagram of the HI in UGC~10214  (inclined feature towards the east) and VV~29c
  (compact feature at $\sim$ 9750~\kms), taken along the major
  axis of UGC~10214, with a slice width of 8\arcsec\ and a position angle of 80~degrees.
  Negative offsets are to the east and positive offsets to the west.}
  \label{fig:hi_pv_diagram}
\end{figure}

These data include both the emission from UGC\,10214 and from the background dwarf galaxy VV\,29c. The velocities of the HI emission from VV\,29c are offset from the velocities of UGC\,10214 at this position, so that  both objects can be clearly distinguished in a position-velocity diagram (Fig.~\ref{fig:hi_pv_diagram}). We separated the emission from UGC\,10214 and VV\,29c into individual data cubes by  masking the channels with velocities between 9630 and 9800\,\kms\ that were in the area of VV\,29c (R.A. between 16:05:3.9 and 16:05:58.3 and Dec between 55:25:00 and 55:25:52). The resulting data cube represents the HI emission from  UGC\,10214 alone. The emission that was masked out in the first step provides the emission of VV\,29c.

We derive the total flux for UGC\,10214 and VV\,29c by summing up all channels from the corresponding data cube over the entire region of each galaxy. The resulting spectrum is then integrated over velocity in the range where the line is visible (8850 --  9800\,\kms\ for UGC\,10214 and 9630 and 9800\,\kms\ for VV\,29c). The error is calculated by the quadratic sum of the error due to the per-channel rms noise and of the calibration error of $\sim$ 3\%\footnote{https://science.nrao.edu/facilities/vla/docs/manuals/oss2013A\break/performance/cal-fluxscale}.

We calculate the atomic gas mass from the total flux, \shi, as:

\begin{equation}
\left(\frac{M_{\rm HI}}{M_{\odot}}\right) =  1.36 \cdot 2.36 \cdot 10^5 \cdot \left(\frac{S_{\rm HI}}{\rm Jy\cdot km s^{-1}}\right) \cdot \left(\frac{D}{\rm Mpc}\right)^2 \quad ,
\end{equation}

where the factor 2.36 is the standard value to calculate the atomic gas mass \citep[e.g.,][]{roberts94, rohlfs96} and the factor 1.36 takes into account the contribution from helium  in order to be consistent with the molecular gas mass calculation.

\section{Methodology}

\subsection{Aperture photometry}
\label{sec:apeture}

Before carrying out the aperture photometry we set all the images, except the SPIRE images and the radio data, to the same spatial resolution and pixel size. We choose the resolution (6\arcsec, corresponding to 4 kpc at the distance of UGC~10214) and the pixel size (2\farcs{45}) of the  MIPS 24\,\mi\ image as reference.  We convert the other images to the MIPS 24~\mi\ point-spread-function using the set of kernels provided by \citet{aniano11}\footnote{http://www.astro.princeton.edu/$\sim$ganiano/Kernels.html}. Then we regrid the images to the MIPS 24\,\mi\ pixel size using the IRAF task \textsc{wregister}, and match at the same time the field of view and orientation. The SPIRE images and the VLA radio data have a coarser resolution than the MIPS 24\,\mi\ map. We therefore exclude these images from any spatially resolved analysis and only measure the total fluxes. The CO CARMA map has a spatial resolution (6\farcs{23} $\times$ 7\farcs{23}), which is  slightly larger than that of the MIPS 24\,\mi\ map. We neglect this small difference and use the regridded (but unconvolved) CO map for our analysis.

We carry out aperture photometry within an ellipse encompassing the entire galaxy disk at all wavelengths in order to obtain the total fluxes. For most images, except \halpha\ and CARMA, this ellipse has a major semi-axis of 80\arcsec and a minor semi-axis of 55\arcsec. The size was chosen to be large enough to cover the complete emission of the disk at all wavelengths. For \halpha, we use a smaller aperture (52\farcs{5} $\times$ 32\farcs{0}) which includes the entire \halpha\ emission but is small enough to avoid a region to the west with some weak artefacts.

In addition, we perform aperture photometry for three regions in the disk, characterised by different star formation properties (see Fig.~\ref{fig:insets} for the contour of the apertures) for their comparison. We separate the regions based on the distribution of FUV and \halpha\ emissions: regions 1 and 3 contain the maxima in the FUV, but only weak \halpha\ emission, whereas region 2 in the centre shows the opposite behaviour (the SPIRE and VLA data were excluded from this analysis). The total area covered by the 3 regions is smaller than the elliptical aperture used for the measurement of the  emission of the entire disk. 

We apply, following the IRAC handbook, aperture correction factors to the \spitzer IRAC images, with the values of  0.92,  0.94, 0.80 and 0.77 for the IRAC bands 1 to 4  band for the entire galaxies and 0.93,  0.95, 0.85 and 0.80 for the IRAC bands 1 to 4  band for the smaller subregions.

All fluxes are corrected for Galactic extinction adopting $A_{\rm V}$ = 0.025 mag from the {\it Nasa Extragalactic Database} which was calculated following the recalibration of \citet{schlafly11} of the \citet{schlegel98} dust maps. The recalibration to other wavelengths was done following the \citet{cardelli89} extinction law\footnote{We used the absorption law calculator by Dough Welch on http://www.dougwelch.org/Acurve.html}.

The photometry is performed using different options of the Python's Astropy package\footnote{\citealt{2013A&A...558A..33A}, \citealt{2018arXiv180102634T}.}. In particular, we use the affiliated package Photutils\footnote{\citealt{bradley17}.} to carry out the photometry with the elliptical apertures. For the photometry of regions 1--3, we define irregular polygonal masks with the package Astropy for each region and determine the total fluxes within the mask using the NumPy package \citep{:/content/aip/journal/cise/13/2/10.1109/MCSE.2011.37}. The background is measured using the median value of the pixels surrounding the galaxy disk which are not covered by the polygonal masks. The resulting fluxes are listed in Table \ref{tab:fluxes}. The errors for the optical and infrared images are determined using the following expression:\

\begin{equation}
\Delta F = \sqrt{F^2\frac{\Delta K^2}{K^2}+\sigma^2 N_A+\sigma^2\frac{N_A^2}{N_S}} \quad ,
\label{eq:flux_err}
\end{equation}

where $K$ and $\Delta K$ are the calibration for each image and its error, $F$ is the flux measured within the aperture, $\sigma$ is the root-mean-square noise within the background aperture, and $N_A$ and $N_S$ are the numbers of pixels in the galaxy aperture and in the aperture used to measure the background emission, respectively. The first term of the sum gives the error associated to the instrument's calibration, the second  is  the error related to the aperture employed, and the third term is the error due to the background subtraction in each image.

The error for the photometry done with the CARMA image, for which the background is zero, and with the FUV map, which is already background-subtracted, was derived from the first two terms in the quadratic sum only.

\begin{table*}
\caption{Fluxes for the total disk and the three disk regions$\tablefootmark{b}$.}           
\label{tab:fluxes}      
\centering          
\begin{tabular}{lccccc}
\hline
\hline
Telescope/Instrument & Unit  & Total  & Flux Reg 1 & Flux Reg 2& Flux Reg 3 \\
Band & & & & &  \\
\hline
GALEX FUV & mJy & $0.29 \pm 0.03$ & $0.073 \pm 0.007$ & $0.072 \pm 0.007$ & $0.098 \pm 0.010$ \\
GALEX NUV & mJy & $0.53 \pm 0.05$ & $0.13 \pm 0.01$ & $0.13 \pm 0.01$ & $0.17 \pm 0.02$ \\ \hline
Sloan u' & mJy & $1.57 \pm 0.04$  & $0.44 \pm 0.01$ & $0.54 \pm 0.01$ & $0.52 \pm 0.01$ \\
Sloan g' & mJy & $7.10 \pm 0.08$  & $1.93 \pm 0.03$ & $2.71 \pm 0.03$ & $1.99 \pm 0.03$ \\
Sloan r' & mJy & $13.63 \pm 0.15$ & $3.60 \pm 0.05$ & $5.81 \pm 0.06$ & $3.45 \pm 0.04$ \\
Sloan i' & mJy & $19.2 \pm 0.2$ &  $5.01 \pm 0.06$ & $8.66 \pm 0.09$ & $4.67 \pm 0.06$ \\
Sloan z' & mJy & $24.22 \pm 0.25$ f & $6.23 \pm 0.07$ & $11.6 \pm 0.1$ & $5.78 \pm 0.07$ \\ \hline
VATT \halpha\ & $10^{-14}$ erg cm$^{-2}$  s$^{-1}$ & $35.8 \pm 1.1$  & $8.2 \pm 0.3$ & $15.0 \pm 0.5$ & $10.9 \pm 0.3$ \\ \hline
IRAC 3.6 & mJy & $18.9 \pm 1.9$ & $4.7 \pm 0.5$ & $9.5 \pm 0.9$ & $4.1 \pm 0.4$ \\
IRAC 4.5 & mJy & $12.2 \pm 1.2$ & $3.1 \pm 0.3$ & $6.0 \pm 0.6$ & $2.8 \pm 0.3$ \\
IRAC 5.8 & mJy & $14.4 \pm 1.4$ & $4.3 \pm 0.4$ & $6.8 \pm 0.7$ & $4.4 \pm 0.4$ \\
IRAC 8.0 & mJy & $26.3 \pm 2.6$ & $8.2 \pm 0.8$ & $10.0 \pm 1.0$ & $9.3 \pm 0.9$ \\ \hline
MIPS 24 & mJy & $23.9 \pm 2.4$ & $6.5 \pm 0.7$ & $7.3 \pm 0.7$ & $9.1 \pm 0.9$ \\ \hline
SPIRE 250 & mJy & $940 \pm 170$ & - & - & - \\
SPIRE 350 & mJy & $450 \pm 100$ & - & - & - \\
SPIRE 500 & mJy & $156 \pm 40$ & - & - & - \\ \hline
CARMA CO & Jy  \kms & $32 \pm 3$ & $10 \pm 1$ & $13 \pm 1$ & $9 \pm 1$ \\ \hline
VLA HI & Jy \kms & 3.8$\pm$0.1 & - & - & - \\
\hline
\end{tabular}
\tablefoot{
\tablefoottext{a} {The apertures used for the regions are described in Sect.~\ref{sec:apeture}  and shown in Fig.~\ref{fig:insets}.}
}
\end{table*}

\subsection{SED modelling with CIGALE}
\label{cigale}

\begin{table*}
\caption{Modeled values (see text) of the different parameters used for the SED modelling with CIGALE.}           
\label{tab:cigale_parameters}      
\centering          
\begin{tabular}{lll}
\hline
\hline
Parameters$\tablefootmark{a}$ & Range of modelled values & Best-fit values$\tablefootmark{b}$\\
\hline
Age of the main population; $t_1$ & 12 Gyr  & 12 Gyr  \\
Age of the last episode of star formation; $t_2$ & 10 -- 1000 Myr & $11\pm 20$ Myr  \\
Timescale of the main population; $\tau_1$ & 2 -- 10 Gyr    & $2.20\pm 0.06$ Gyr\\
Timescale of the last episode of star formation; $\tau_2$ & 10 -- 1000 Myr   & $ 310\pm 340$ Myr \\
Burst fraction; log f & $-4$ to $-1$ &  $-3.7\pm -3.8$\\
IMF & Chabrier & Chabrier \\
Metallicity; Z & 0.02, 0.05  & 0.02 \\
Ionisation parameter; log U & $-3$ &  $-3$ \\
Reddening of populations younger than 10 Myr; $E(B-V)_{\rm young}$ & 0.05 -- 0.70 mag & $0.18\pm0.03$\\
Relative reddening of populations older than 10 Myr; $E(B-V)_{\rm old\ factor}$ & 0.25 -- 0.75 & $0.05\pm0.01$ \\
UV bump amplitude &  0.0, 1.5, 3.0 & 0  \\
Incident dust radiation field intensity powerlaw distribution; $\alpha$ & 1.0 -- 4.0 &  3.0\\
\hline
\end{tabular}
\tablefoot{
\tablefoottext{a} {See text (Sect.~\ref{cigale}) for a detailed explanation of the parameters.}
\tablefoottext{b} {Best-fit values derived from the total fluxes from Table~\ref{tab:fluxes}. The results for the individual regions are similar.}
}
\end{table*}

The SED modelling is carried out with CIGALE \citep[][Boquien et al., in press]{noll09}. CIGALE is based on an energy balance approach, the energy absorbed by dust at short wavelengths is re-emitted self-consistently. The physical properties and the uncertainties are estimated from the likelihood-weighted means and standard deviations over a grid of models.

The code is highly modular to allow for a flexible modeling. In the present case the star formation history is modeled through two decaying exponentials, the first one modeling the long term star formation and the second one the latest episode of star formation. Each exponential has an independent age ($t_1$ and $t_2$) and timescale ($\tau_1$ and $\tau_2$), and are linked through the burst fraction $f$, giving the total fraction of stars formed in the second exponential relative to the total mass of stars ever formed. The stellar emission is computed using the model by \citet{bruzual03}. The nebular emission, which includes recombination lines as well as continuum originating from free--free, free--bound, and 2--photon emission is based on the number of ionising photons\footnote{It is assumed that all ionising photons ionise the gas.} and templates improved from \citet{inoue11} computed from \textsc{CLOUDY} 08.00 \citep{ferland98}. The attenuation is based on a power-law modified starburst curve \citep{calzetti00}, allowing for a differential reddening between stars older and younger than 10~Myr and with an optional UV bump at 217.5~nm. Finally, dust emission is modeled using the \citet{dale14} templates. In the unique case of determining the dust mass we only fitted infrared data with the dust templates used by \citet{draine14}. This yields a total of 20\,003\,760 models. 

The run is based on a pre-release of version 0.12 modified to fit the \halpha\ emission in addition to broadband fluxes.

\begin{figure}[h!]
  \centering
    \includegraphics[width=0.5\textwidth]{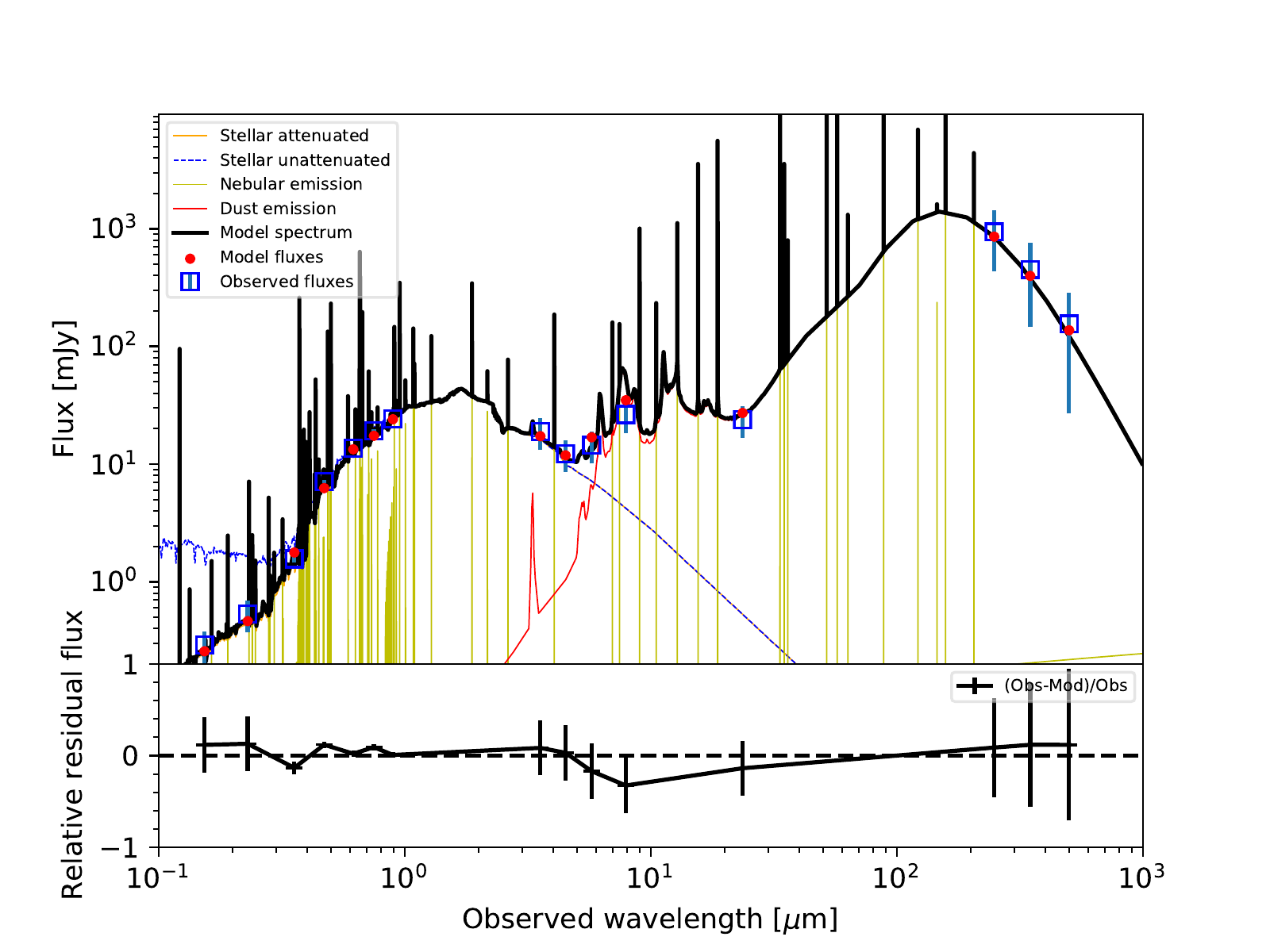}
  \caption{SED fitting with CIGALE for the disk of UGC~10214.}
  \label{fig:cigale_SED}
\end{figure}

The range of the modeled values used for the different parameters as well as the best-fit parameters for the total galaxy are listed in Table~\ref{tab:cigale_parameters} (the best-fit values for the individual region are very similar). The beginning of the long term star formation (parameter $t_1$) was fixed to a reasonable value of 12 Gyr in order to reduce the number of free parameters. The metallicities were tested for two  values (roughly solar and super-solar) that are plausible for this high mass galaxy and for which stellar emission models were available from  \citet{bruzual03}. No satisfactory fit was achieved with the super-solar metallicity. The best-fit results for the last  episode of SF give a time-scale of several 100\,Myr and an age of  several 10\,Myr. The stellar mass fraction formed in this last episode is small (0.02\%) which is not surprising given the large stellar mass of UGC~10214. Table~\ref{tab:cigale_results} gives the SFRs averaged over 10 and 100\,Myr and the stellar masses for the total disk and the individual regions, as well as the total dust masses which will be discussed in more detail in Sect.~4. The best-fit SED for the total galaxy is shown in Fig.~{\ref{fig:cigale_SED}}. 

\begin{table*}
\caption{Results from the modelling with CIGALE.}
\label{tab:cigale_results}      
\centering          
\begin{tabular}{l c c c c}
\hline
\hline
Region & SFR(10 Myr)  & SFR (100 Myr) & \mstar &\mdust\\
& [\msun\  yr$^{-1}$] & [\msun\ yr$^{-1}$]  & [10$^{10}$ \msun]  & [10$^{8}$ \msun]\\
\hline
Disk & $4.8 \pm 0.5$ & $0.77 \pm 0.08$ & $12.8 \pm 0.7$ & $1.7\pm0.3$\\
Region 1 & $1.31 \pm 0.09$ & $0.21 \pm 0.03$ & $3.4 \pm 0.2$ &-\\
Region 2 & $1.29 \pm 0.06$ & $0.26\pm 0.06$ & $5.9 \pm 0.3$ &-\\
Region 3 & $1.4 \pm 0.2$ & $0.4 \pm 0.2$ & $3.1 \pm 0.5$ & - \\     
\hline         
\end{tabular}

\end{table*}

\subsection{Pixel-by-pixel analysis}

Apart from analysing the integrated fluxes for the entire galaxy and the three regions, we perform a comparison of the SFR, molecular gas mass and stellar mass on a pixel-by-pixel basis, using the FUV, \halpha, CARMA, and IRAC 3.6\,\mi\ maps.

\begin{figure}
\centering
\includegraphics[width=\linewidth]{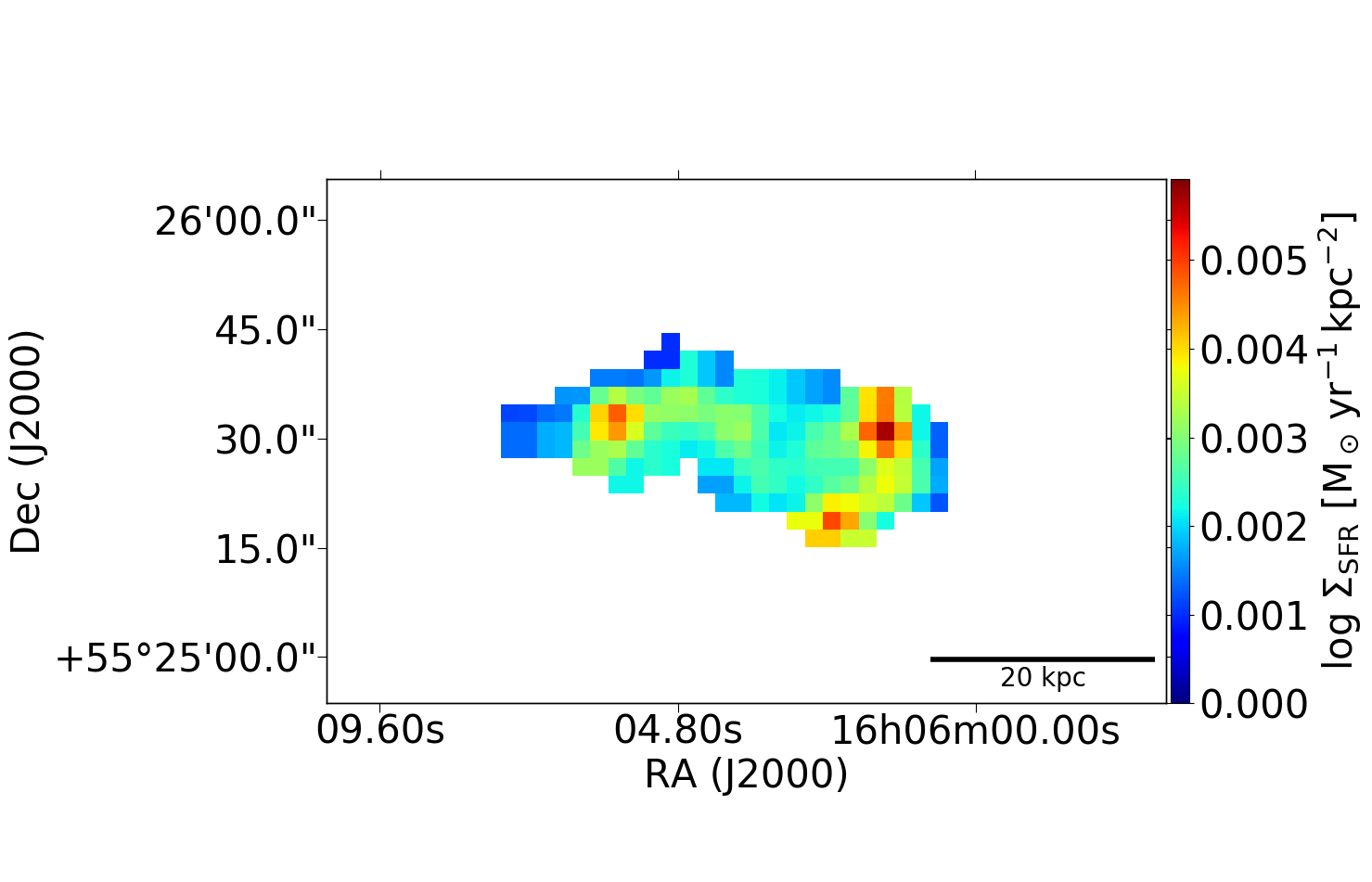}
\caption{Final mask used for the pixel-by-pixel analysis, applied for illustration to the SFR map obtained from the dust-corrected FUV emission 
(see Sect.~\ref{sect:sfr}).}
\label{fig:pixmask}
\end{figure}

We only take into account those pixels for which all five images had significant fluxes at a 3$\sigma$ level. For this, we measure the mean value and standard deviation outside the main galaxy in all five images and create masks consisting of those pixels with fluxes more than three times the standard deviation. The final mask is shown in  Fig.~\ref{fig:pixmask}. The size of the mask is mainly restricted by the CO map.

\section{Results}
\subsection{Atomic and molecular gas and dust}

\subsubsection{Atomic and molecular gas mass and distribution}

The  molecular gas in the CARMA map shows an asymmetric distribution. In the eastern part (region 1), the molecular gas has an elongated distribution with a peak. The molecular gas surface density is highest in this region. In the middle (region 2) and western part (region 3) there is a double-ring structure of lower intensity (see Fig.~\ref{fig:insets}). The western-most ring is reflected in the SFR, clearly seen in the IRAC 8.0\,\mi, MIPS 24\,\mi, and \halpha\ images.

\begin{figure}
  \centering
\includegraphics[width=0.5\textwidth,trim=0.cm 0.cm 0.cm 0.cm,clip]{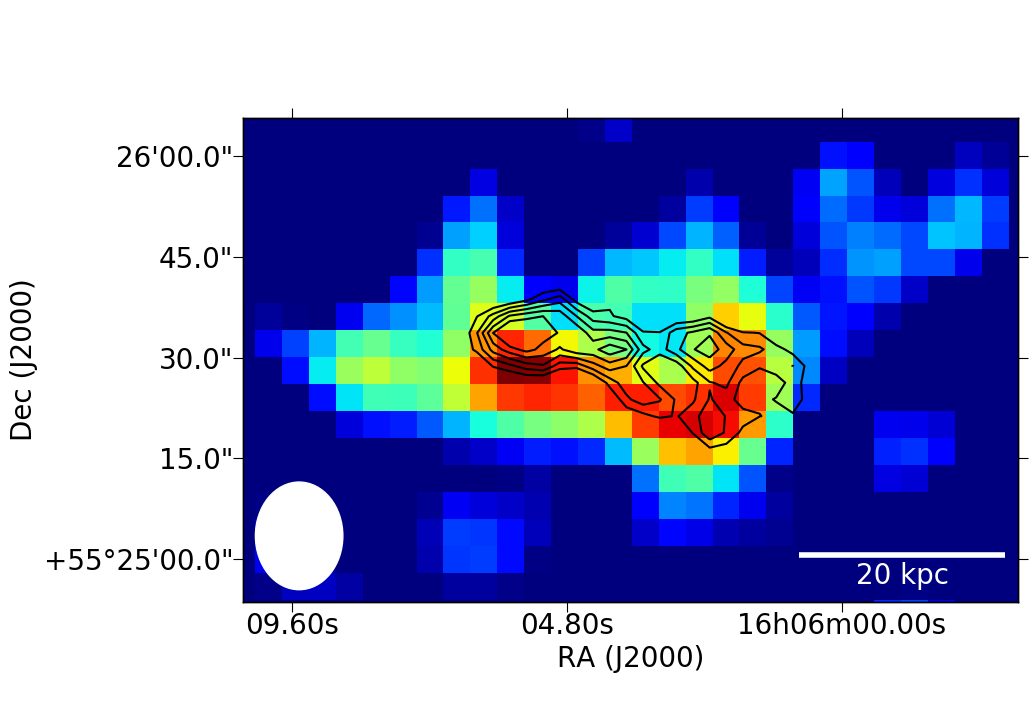}
  \caption{Distribution of the HI emission of UGC 10214 (colour) and the CO(1-0) data from
  CARMA (contours, ranging from 6 to 14\,$\sigma$ in 2\,$\sigma$ steps). The ellipse in the lower left corner shows the
  resolution of the HI data.
  }
    \label{fig:map_hi_carma}
\end{figure}

The atomic gas is much more extended than the molecular gas (Fig.~\ref{fig:map_hi_carma}). It shows a ridge-like structure from east to west which is offset to the south from the molecular gas. In the western end it shows an arc-like structure which coincides with the western end of the molecular gas distribution.

The total atomic gas mass, $\left(2.21\pm0.07\right)\times$10$^{10}$\,\msun, see Tab.~\ref{tab:results}) is a  factor 3.7 higher than the total molecular gas mass from CARMA,$\left(6.0\pm0.6\right)\times$10$^{9}$\,\msun. This factor is slightly lower when using the molecular gas mass from the IRAM single-dish data, $\left(9.0\pm1.5\right)\times$10$^{9}$\,\msun, corresponding to \mhi/\mmol  $\sim$2.5.

\subsubsection{Kinematics of the atomic and molecular gas}

\begin{figure}
  \centering
\includegraphics[width=0.5\textwidth,trim=0.cm 0.cm 0.cm 0.cm,clip]{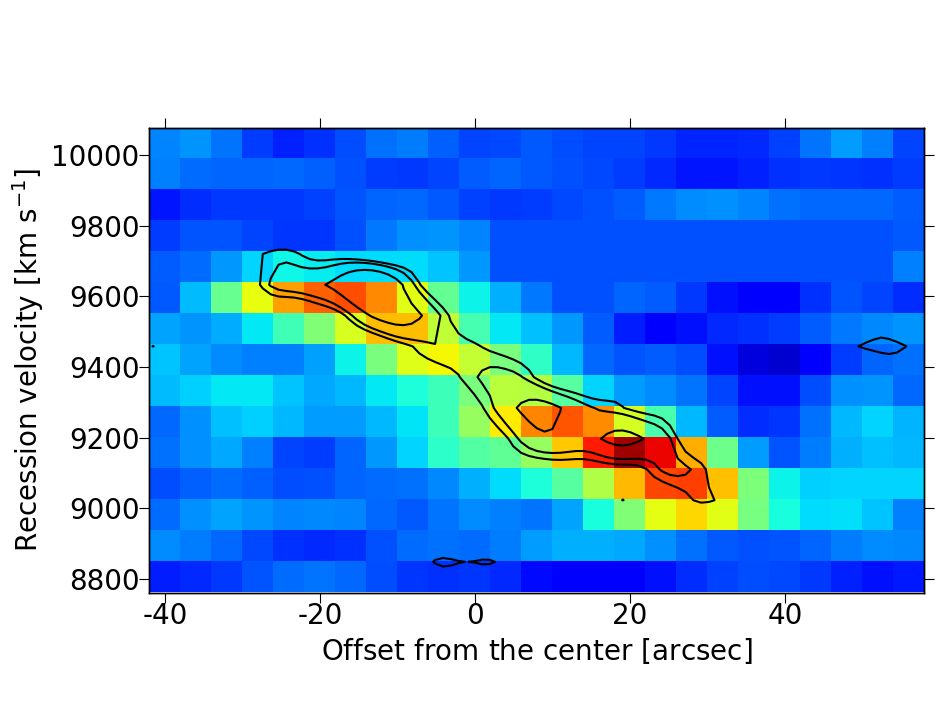}
  \caption{Position-velocity diagram of UGC\,10214 from the HI data (colour) and the CO(1-0) data from
  CARMA (contours), taken along the major axis of UGC\,10214 as in Fig.~\ref{fig:hi_pv_diagram} with a slice width of 4\arcsec on either side. 
  }
    \label{fig:pv_carma_hi}
\end{figure}

Figure~\ref{fig:pv_carma_hi} shows the position-velocity (pv) diagram of HI and of CO (from CARMA) along the major axis of the disk of UGC~10214. Both the HI and CO follow a regular velocity distribution indicating a rotational disk. The HI lines are wider and their distributions are more extended towards the western side than the CO. 

The high velocity resolution of the IRAM observations provides additional information about the kinematics of the gas (Fig.~\ref{fig:iram_co_spectra}). Consistent with the CARMA data, a velocity gradient from the eastern to the western part of the galaxy is visible in the IRAM spectra. In addition, the line-shape changes drastically from a relatively weak single-peak spectrum in the east (position 1), to broader, weaker lines in the middle (positions 2, 3, 4, and 8) to a strong peak with a broader, weaker secondary peak in the west (positions 5, 6, and 7).

\begin{figure}
  \centering
    \includegraphics[width=0.5\textwidth,trim=5.cm 0.cm 0.cm 0.cm,clip]{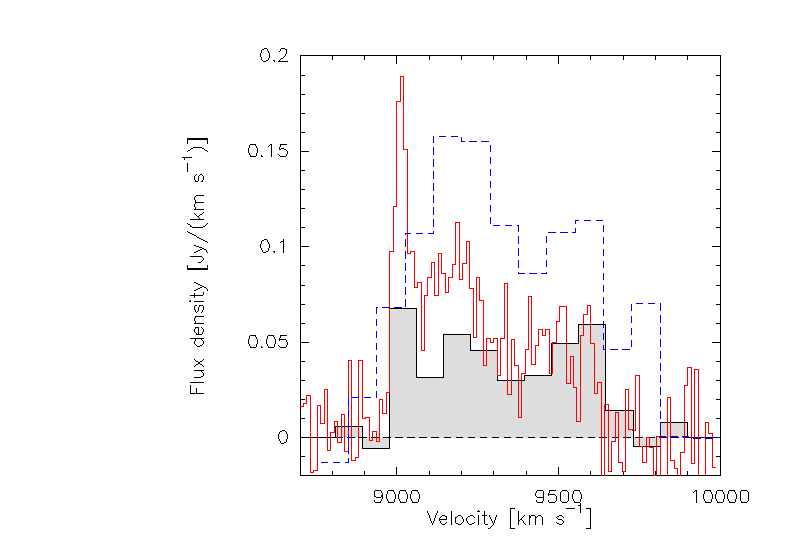}
  \caption{CO(1-0)  spectra  form the IRAM 30m telescope (red, continuous line) and from CARMA (grey shaded), 
  together with the HI emission (blue dashed line)
  integrated over the entire disk of the Tadpole galaxy. The HI spectrum contains only the
  emission of UGC\,10214 (i.e. the emission from VV\,29c has been excluded as described in
  Sect.~\ref{sec:hi_data}). The HI flux is multiplied by a factor
  of 15 for better visibility.
}
  \label{fig:iram_carma_hi_spectra}
\end{figure}

Figure~\ref{fig:iram_carma_hi_spectra} shows the HI and CO spectra, integrated over the entire disk. The HI spectrum extends to slighly higher velocities, consistent with a more extended disk. The HI spectrum follows more closely the IRAM spectrum and includes the peak at 9200~\kms. The strong peak at 9000~\kms\ visible in the IRAM spectrum would be lower by a factor of about 2 when smoothed to the velocity resolution of the CARMA and HI data. Taking this decrease into account, it is only slightly higher than the CARMA data. The HI data, on the other hand, shows no indication for a strong emission at this velocity.

\subsubsection{Comparison between IRAM and CARMA data}
\label{sec:comparison_iram_carma}

The total molecular gas mass obtained with the IRAM measurements is $\left(9 \pm 2\right)\times 10^{9}$\,\msun, a factor of about 1.5 higher than the  total flux from CARMA. This indicates that some flux ($\sim$30\%) might be missing in the interferometer measurements. 
 
In Table~\ref{tab:ico_mmol} we list the ratio between the molecular gas masses calculated from the IRAM velocity integrated intensities (from Eq.~\ref{eq:mh2_iram}) and from the CARMA data. For this comparison we multiply the velocity integrated CARMA map (from Fig.~\ref{fig:insets}) with a Gaussian beam of FWHM 20\arcsec\ (in order to convolve from the CARMA resolution to the IRAM FWHM) for each position observed by IRAM and measure the total flux from the resulting image. We then calculate the total molecular gas mass from Eq.~\ref{eq:mh2_carma}. The ratio of the CARMA-to-IRAM gas masses spans a range from 0.4 to 1.3, with the higher values being present in the eastern part of the galaxy and the lower in the western. This indicates that in the western part some flux might be missing from the interferometer measurements.

We compare the total CO spectra, integrated over the entire disk, measured with both CARMA and with the IRAM 30\,m in Fig.~\ref{fig:iram_carma_hi_spectra}. The total line widths agree well. At $v \sim 9000$\,\kms, there is a pronounced peak in the IRAM spectrum that is not seen in the CARMA spectrum due to its lower resolution. The  missing flux seems to come in part from this peak as well as from the broad velocity feature around 9200\,\kms. Both features are  emitted in the western part of the galaxy (see Fig.~\ref{fig:iram_co_spectra}).

\subsubsection{Comparison of \icoone\ and \icotwo}

To interpret the ratio of \icotwo/\icoone\ one has to consider two main parameters: the source distribution and the opacity. For optically thick, thermalised emission with a point-like distribution we expect a ratio  \icotwo/\icoone\ = $(\theta_{\rm CO(1-0)}/\theta_{\rm CO(2-1)})^2$  = 4 (with $I_{\rm CO}$ in $T_{\rm mB}$ and $\theta$ being the FWHM of the beams). On the other hand, for a uniform source brightness distribution we expect ratios over 1 for optically thin gas, and ratios between about 0.6 and 1 for optically thick gas (with excitation temperatures above 5\,K). Thus, values of \icotwo/\icoone $>$ 1 indicate a source distribution with an extension less than $\theta_{\rm CO(1-0)}$, whereas \icotwo/\icoone $\lesssim$ 0.6 indicates diffuse, subthermally excited gas.

The  mean line ratio, averaged over all pointings is \icotwo/\icoone  = $0.8 \pm 0.2$. This value is close to  the mean value found by \citet{leroy09} from CO(2-1) and CO(1-0) maps for nearby galaxies from the SINGS sample (\icotwo/\icoone $\sim$ 0.8) and those from \citet{braine93} who obtained a mean line ratio of  \icotwo/\icoone = $0.89 \pm 0.06$ for a sample of nearby spiral galaxies. Both values are, in contrast to ours, corrected for beam-size effects. The ratios of the \icotwo/\icoone\ of the individual pointings are between $1.2\pm 0.4$ (position 7) and $< 0.6 \pm 0.2$ (position 5). Although these values are, within the errors, compatible with the above mentioned reference values \citep{leroy09, braine93}, the low value at position 5 could indicate a subthermal excitation and the presence of diffuse gas which might be responsible for the broad emission feature in the spectrum around 9200~\kms\ (Fig.~\ref{fig:iram_co_spectra}).

\subsubsection{Gas-to-dust mass ratio}

The CIGALE fitting of the total IR to submm SED of the disk of UGC~10214 yields a dust mass of \mdust = $\left(1.7 \pm 0.3\right)\times10^8$\,\msun. The total dust-to-gas mass ratio is thus $180 \pm 30$, taking the molecular gas mass from the IRAM measurement. This value is very close to that of the local  Galaxy (\mgas/\mdust = 186, taking the He fraction of 1.36 into account, \citealt[][Tab. 2]{draine07b}), in agreement with the similar metallicity.

\subsubsection{Atomic and molecular gas in VV\,29c}

As described in Sect.~\ref{sec:hi_data}, we separate the total HI datacube into the HI emission of VV~29c from that of UGC~10214. We then derive the total, spatially integrated, HI spectrum from VV\,29c and measure the HI flux by integrating this spectrum over the velocity range  between 9630 and 9800\,\kms. We obtain a total HI flux of $\left(0.60 \pm 0.03\right)$\,Jy\,\kms, and a total HI mass of (3.5$\pm$0.2)$\times 10^9$\,\msun. This value is very close to that found by \citet{briggs01} of 3.1$\times 10^9$\,\msun\ (including Helium and for a Hubble constant of 63\,\kms/Mpc).

We do not detect any CO emission from this region and velocity range (see the pv diagram of CO in Fig.~\ref{fig:pv_carma_hi}). We sum the CO spectra over the same region where HI emission from VV~29c is detected and determine a 3$\sigma$ upper limit of 0.6\,Jy\,\kms, corresponding to a molecular gas mass of $1\times 10^8$\,\msun. The molecular gas fraction, \mmol/\mhi\ $<$ 0.03, is thus much lower than for UGC~10214, and typical for a dwarf galaxy.

\subsection{Star formation rate }
\label{sect:sfr}

Both \halpha\ and FUV are SF tracers, albeit sensitive to different timescales. Whereas \halpha, produced by ionising  stars, traces the current SFR (up to $\sim$10\,My), the FUV emission is emitted mainly by less massive stars with life-times of up to $\sim$100\,Myr. The exact values of these timescales are uncertain and depend strongly on the SF history (see Sect.~\ref{sec:sf_history}  for a discussion). In the following we refer to the timescale traced by the FUV as the ``recent'' SFR, as opposed to the ``current'' SFR traced by \halpha.  Both tracers are very sensitive to dust extinction. We therefore use hybrid tracers, combining them with the dust re-emission at 24\,\mi, in order to also take into account the dust-obscured SF.

We use the prescriptions in Tables 1 and 2 of \citet*{kennicutt12} \citep[taken from][]{kennicutt09,hao11,murphy11} to derive SFRs, giving:

\begin{equation}
\log {\rm SFR}_{\rm H\alpha}\left[ M{_\odot} \rm{yr}^{-1}\right]=\log (L({\rm H\alpha})+0.020 L\left({\rm 24\,\mu m})\right)-41.27 ,
\label{eq:sfr_halpha}
\end{equation}

\begin{equation}
\log {\rm SFR}_{\rm FUV}\left[ M{_\odot} \rm{yr}^{-1}\right]=\log (L({\rm FUV})+3.89 L\left({\rm 24\,\mu m})\right)-43.35 ,
\label{eq:sfr_uv}
\end{equation}

where the luminosities are in units of erg\,s$^{-1}$ and $L({\rm 24\,\mu m})$ and $L({\rm FUV})$ are calculated from the fluxes $F_{\rm \lambda}$ as $L({\rm \lambda}) = 4\pi D^2 \lambda F_{\rm \lambda}$. We assume, following  \citet{hao11}, that  \textit{IRAS} 25\,\mi\ and MIPS 24\,\mi\ luminosities can be used interchangeably for this estimation. The SFRs are calculated for a Kroupa Initial Mass Function \citep[IMF,][]{kroupa01} which is very similar to the Chabrier IMF used in the CIGALE modelling.
 
\begin{figure}[h!]
  \centering
  \begin{subfigure}[b]{0.5\textwidth}
    \includegraphics[width=\textwidth]{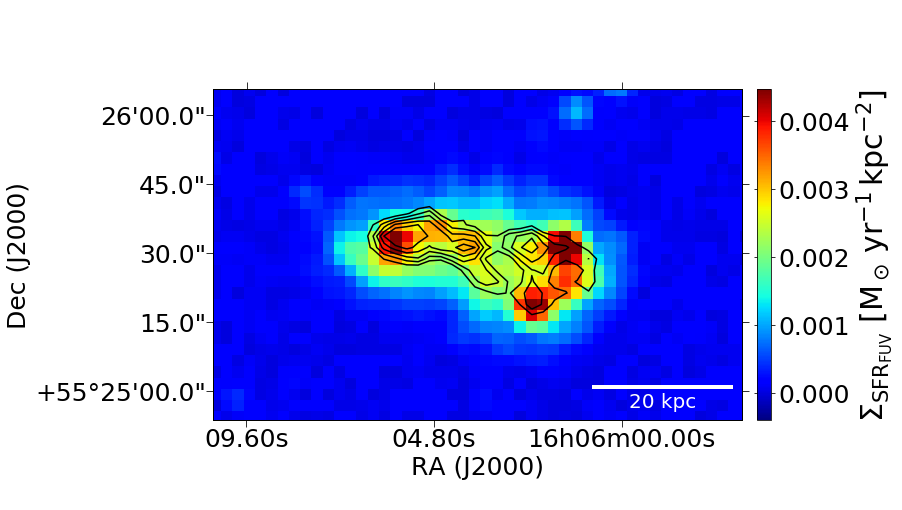}
    \caption{}
  \end{subfigure}
  \begin{subfigure}[b]{0.5\textwidth}
    \includegraphics[width=\textwidth]{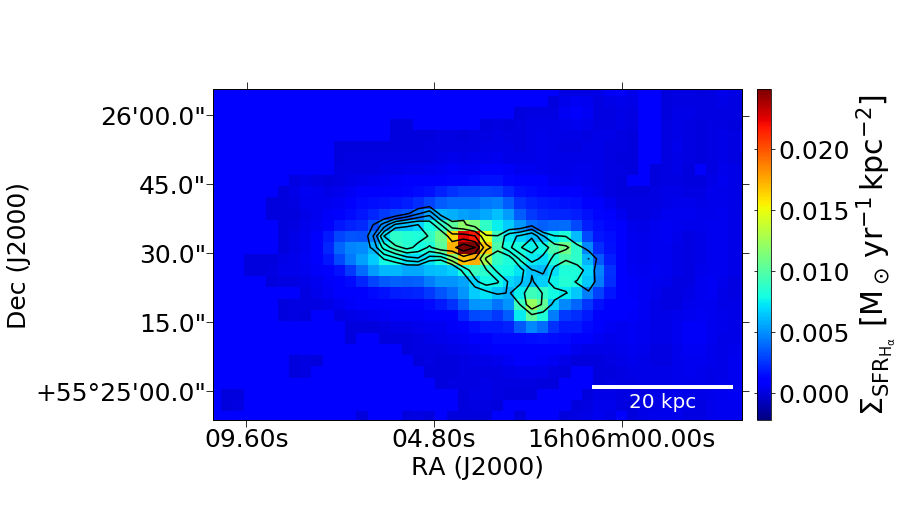}
    \caption{}
  \end{subfigure}
  \caption{CO contour levels (ranging from 6 to 14\,$\sigma$ in 2\,$\sigma$ steps) over the maps of (a) \sfrfuv\  and (b)  \sfrhalpha.
  }
  \label{fig:overlay}
\end{figure}

With these prescriptions, we  produce maps for the SFR. Fig.~\ref{fig:overlay} shows the resulting maps, overlaid with contours of the CARMA CO image. A striking result is the different distributions of the SFRs based on \halpha\ and FUV. Whereas  \sfrfuv\ shows peaks in the east and along the western spiral arm, \sfrhalpha\ has a pronounced maximum in the centre. The ridge towards the east and the western spiral arm are also visible in \sfrhalpha\  but are fainter than the centre. The eastern ridge of \sfrfuv\ is closely  followed by the CO emission which extends until the peak of \sfrhalpha\ in the centre. The western part  of the spiral arm is also followed by the CO and also by HI (see Fig.~\ref{fig:map_hi_carma}).
 
Finally, based on the fluxes from Table~\ref{tab:fluxes} we calculate the SFR for each region as well as for the entire disk which are listed in Tab.~\ref{tab:results}. The SFR from the combination of \halpha+24\,\mi\ (\sfrhalpha) is always higher (by a factor of about 2 -- 4) than the SFR derived from FUV+24\mi\ (\sfrfuv). The difference is more pronounced in the centre (region 2). This  trend is also seen in the modelling results with CIGALE (Tab.~\ref{tab:cigale_results}) where the values for the current SFR (SFR$_{10\rm{Myr}}$ which corresponds to the time-scale of \halpha) are considerably higher (factor $3.5-6$) than for the SFR in the past 100\,Myr, SFR$_{100\rm{Myr}}$. The values for \sfrhalpha\ and SFR$_{10\rm{Myr}}$ generally agree very well (expect for region 1), but the values for \sfrfuv\ are a factor of $1.5 - 2$ higher than SFR$_{100\rm{Myr}}$. A reason for this discrepancy might be that the time-scale for the FUV traced SFR is on the order of several tens to 100\,Myr (see Sect.~\ref{sec:sf_history}), so that \sfrfuv\  traces a somewhat shorter time than SFR$_{100\rm{Myr}}$. In the case of a decreasing SFR,  as modelled here, the SFR averaged  over a shorter timescale results in a higher value. In any case, both methods for deriving the SF history indicates that the SFR has been increasing in the recent past.

In Table~\ref{tab:results} we also list the molecular gas mass (from Eq.~\ref{eq:mh2_carma}) for the different regions and the resulting molecular gas depletion time, \taudep = \mmol/SFR. The current molecular gas depletion time (from \sfrhalpha) is very close to the value found by \citet{bigiel11}, and the molecular gas depletion time from \sfrfuv\  is a factor $\sim$2 lower.  The variations of \taudep\ between the different regions are less than a factor 2 with a trend of having a shorter \taudep\ in the western side (region 3).

\begin{table*}
\caption{Values of SFR, \mstar, \mmol, sSFR, and molecular gas depletion time estimated for the disk and the three disk regions.}
\label{tab:results}
\centering
\begin{tabular}{l c c c c}
\hline
\hline
Band & Total & Reg 1 & Reg 2 & Reg 3 \\
\hline
\sfrfuv\ [\msun\ yr$^{-1}$ ] & $1.66 \pm 0.17$ & $0.44 \pm 0.04$ & $0.47 \pm 0.05$ & $0.61 \pm 0.06$ \\
\sfrhalpha\ [\msun\ yr$^{-1}$ ] & $4.73 \pm 0.19$ & $1.11 \pm 0.05$ & $1.90 \pm 0.07$ & $1.49 \pm 0.06$ \\
\hline
\mstar\ [$10^9$ \msun ] &  $128 \pm 13$ & $32 \pm 3$ & $64 \pm 6$ & $28 \pm 3$ \\
\mmol\ [$10^9$ \msun ] (CARMA) & $6.0 \pm 0.6$ & $1.9 \pm 0.2$ & $2.4 \pm 0.2$ & $1.7 \pm 0.2$ \\
\mmol\ [$10^9$ \msun ] (IRAM) & $9.0 \pm 1.5$ &  &  &  \\
\mhi\  [$10^9$ \msun ] & $22.1\pm$0.7  & - & - & -\\
\hline
sSFR (FUV) [10$^{-12}$ yr$^{-1}$] & $13 \pm 3$ & $14 \pm 3$ & $7.3 \pm 1.5$ & $22 \pm 4$ \\
sSFR (H$\alpha$) [10$^{-12}$ yr$^{-1}$] & $36 \pm 5$ & $35 \pm 5$ & $30 \pm 4$ & $53 \pm 8$ \\
\hline
\taudep\tablefootmark{a} (FUV) [Gyr]  (CARMA) &$3.6 \pm 0.7$ & $4.4 \pm 0.9$ & $5.1 \pm 1.0$ & $2.8 \pm 0.6$ \\
\taudep\tablefootmark{a}(FUV) [Gyr]  (IRAM) & $5.4 \pm 1.4$ & & & \\
\taudep\tablefootmark{a}(\halpha) [Gyr]  (CARMA) & $1.3\pm 0.2$ & $1.75 \pm 0.25$ & $1.3 \pm 0.2$ & $1.1 \pm 0.2$ \\
\taudep\tablefootmark{a} (\halpha) [Gyr]  (IRAM) & $1.9 \pm 0.4$ & && \\
\hline
\end{tabular}
\tablefoot{
\tablefoottext{a} {Molecular gas depletion time \taudep = SFR/\mmol .}
}
\end{table*}

\subsection{Stellar mass}

The  IRAC 3.6\,\mi\ emission is a good tracer of the low-mass stars which dominate the total stellar mass. We therefore use the 3.6~\mi\ emission as a tracer for the stellar mass, following as a starting point the calibration by \citet{oliver10} for the morphological type of UGC\,10214 (Sb(s)c, according to the \textit{NASA Extragalactic Database}),  \mstar [\msun] = 27.6 L({3.6\,\mi)}[\lsun]. This calibration is based on a Salpeter IMF, and we apply a correction factor of 0.626 to convert this expression to a Chabrier IMF and to achieve agreement with the stellar mass from CIGALE for the entire disk. Thus, we calculate the stellar mass as: 

\begin{equation}
\frac{M_\star}{M_{\odot}}=14.5\times\frac{L({3.6\,\mu \rm{m})}}{L_\odot} \quad .
\label{eq:mstar}
\end{equation}

Based on the fluxes from Table~\ref{tab:fluxes} we calculate the stellar mass for each region as well as for the entire disk (Table~\ref{tab:results}). These values for \mstar\ for region 1--3 agree  within the errors with those calculated with CIGALE (Table~\ref{tab:cigale_results})   showing that the simple  prescription of Eq.~\ref{eq:mstar} is a good approximation. In the central region of the galaxy (region 2), the stellar mass is higher and the specific SFR, sSFR = SFR/\mstar, is lower, in particular when considering \sfrfuv. The highest values of sSFR are in the western part, both for \sfrfuv\ and for \sfrhalpha.

\subsection{Pixel-by-pixel analysis of star formation and gas}
\subsubsection{The Kennicutt-Schmidt law}

\begin{figure*}[h!]
  \centering
    \includegraphics[width=\textwidth]{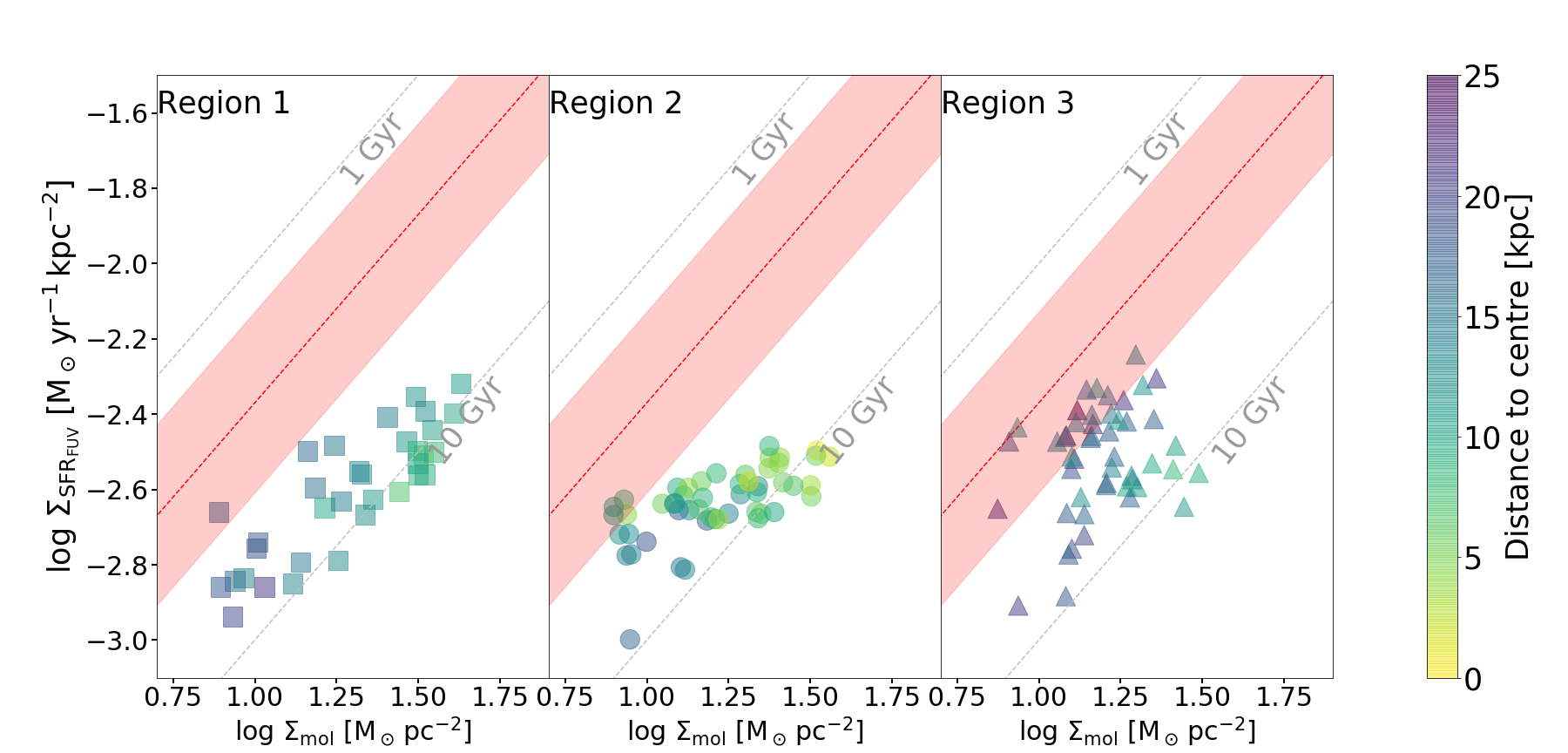}
    \includegraphics[width=\textwidth]{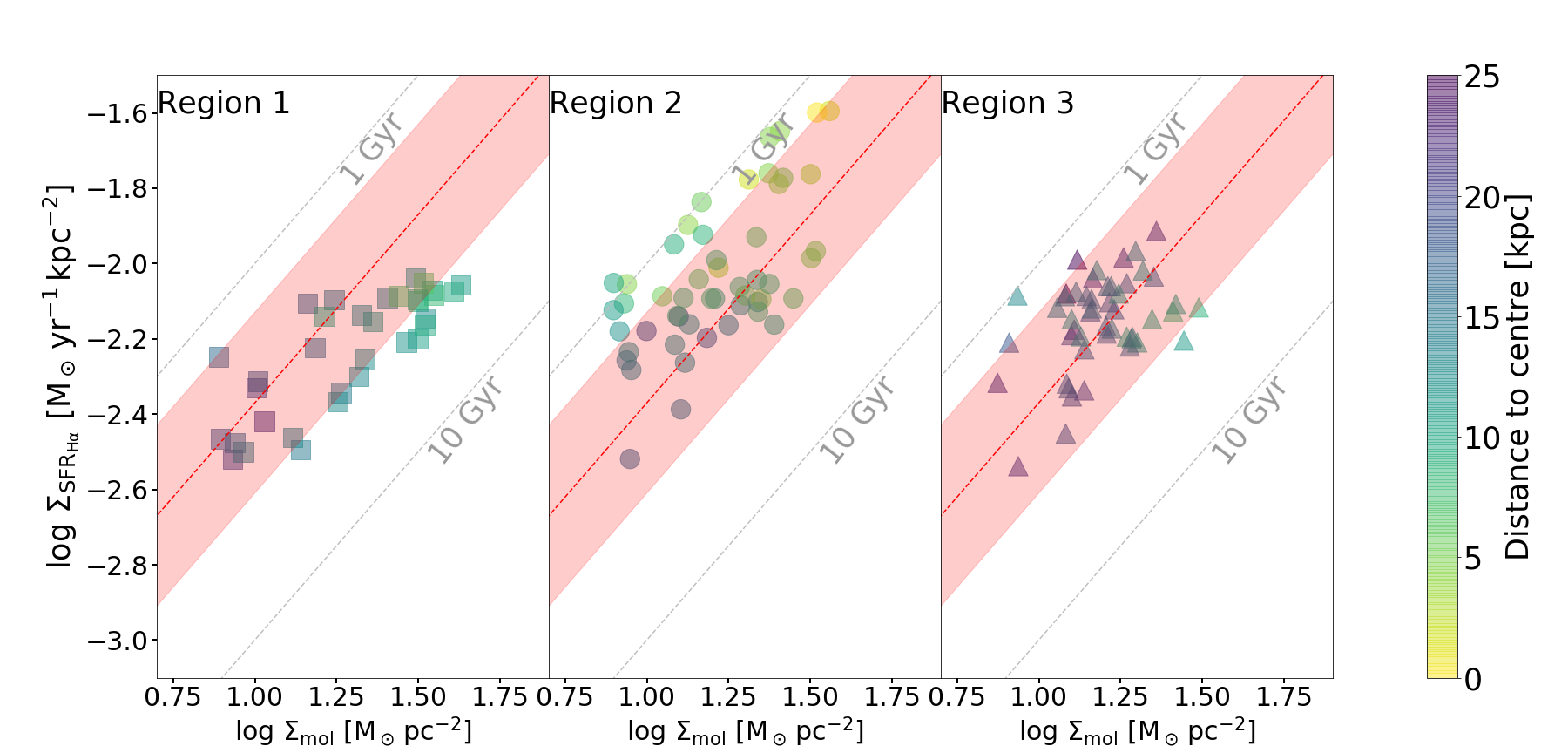}
  \caption{SFR per area versus molecular gas surface density (Kennicutt-Schmidt relation) for different regions in the disk of UGC~10214. The upper row shows the SFR derived from FUV and 24~\mi\ emission, \sfrfuv, and the lower row the SFR derived from the \halpha\ and 24~\mi\ emissions, \sfrhalpha. The pixels in the separate panels correspond to the different regions shown in Fig.~\ref{fig:insets}. The dashed grey lines represent different molecular gas depletion times. The dashed red line represents the depletion time estimated for star-forming galaxies  \citep[2.35\,Gyr,][]{bigiel11} and its dispersion (0.23dex), represented by the pink-shaded area. The colour of the markers of the different regions indicates the distance to the galaxy centre.}
  \label{fig:KSpix}
\end{figure*}

Figure~\ref{fig:KSpix} shows the SFR per area as a function of the surface density of the molecular gas (the Kennicutt-Schmidt relation), for the individual pixels in the images. We show the relation for both \sfrhalpha\ and \sfrfuv\ for region 1, region 2, and region 3 separately. We furthermore indicate the distance from the centre as the colour coding.
 
The most striking result is that \sigmasfrhalpha\ is systematically higher than \sigmasfrfuv\ for a given molecular mass surface density. This effect has already been seen in the global and regional analysis (Sect.~\ref{sect:sfr}) and indicates a higher current SFR, traced by \halpha, compared to the  recent SFR,  traced by FUV. The pixel-to-pixel analysis shows that this trend is present over the entire disk, and that it is particularly pronounced in region 2 (the central region) where the molecular depletion time, \taudep, is about 0.8--0.9\,dex lower for  \sfrhalpha\ than for \sfrfuv. The difference in the other regions is less pronounced ($\sim$0.4\,dex). 

In regions 1 and 3, the relation between \sigmasfr\ and \sigmamol\ have similar shapes for the FUV and \halpha\ traced SFRs, the only difference being the systematically higher values for \sigmasfrhalpha. This means that the increase in the current SFR has taken place  over the entire galaxy. The mean molecular gas depletion times are slightly lower (by 0.2\,dex) in region 3 than in region 1, both for \sfrhalpha\ and for \sfrfuv. In region 2, the situation is different. Whereas \sigmasfrhalpha\  shows a roughly linear increase with \sigmamol,  the  relation of \sigmasfrfuv\  with \sigmamol\ is flat which means that the molecular depletion time, \taudepfuv, changes systematically as a function of \sigmamol\ and  as a function of distance to the centre. Whereas  at large central distance,  \taudepfuv\ is similar to the values in region 1 and 3, closer to the centre \taudepfuv $\approx$ 10\,Gyr. 

The mean values for the molecular depletion times, averaged over all pixels, are \taudepfuv = 6.9\,Gyr (with a standard deviation of 2.5 Gyr)  and \taudephalpha = 2.4\,Gyr (with a standard deviation of 1.0\,Gyr). The depletion time from \halpha\ is very similiar to the result of \citet{bigiel11} from a spatially resolved analysis of 30 spiral galaxies based on FUV+24\mi\ as a SFR tracer (\taudep = 2.35\,Gyr with a dispersion of 0.24\,dex), whereas \taudepfuv\ is a factor of 2.9 longer  (0.46\,dex below). Possibly, the long \taudepfuv\ reflects the changes in the SF history rather than indicating a low star formation efficiency (SFE, the inverse of the gas depletion time) because the molecular gas that we observe is related to the current SFR (traced by \halpha) and not to the SF several tens of Myr ago.

\subsubsection{SFR and stellar mass}

\begin{figure*}[h!]
  \centering
    \includegraphics[width=\textwidth]{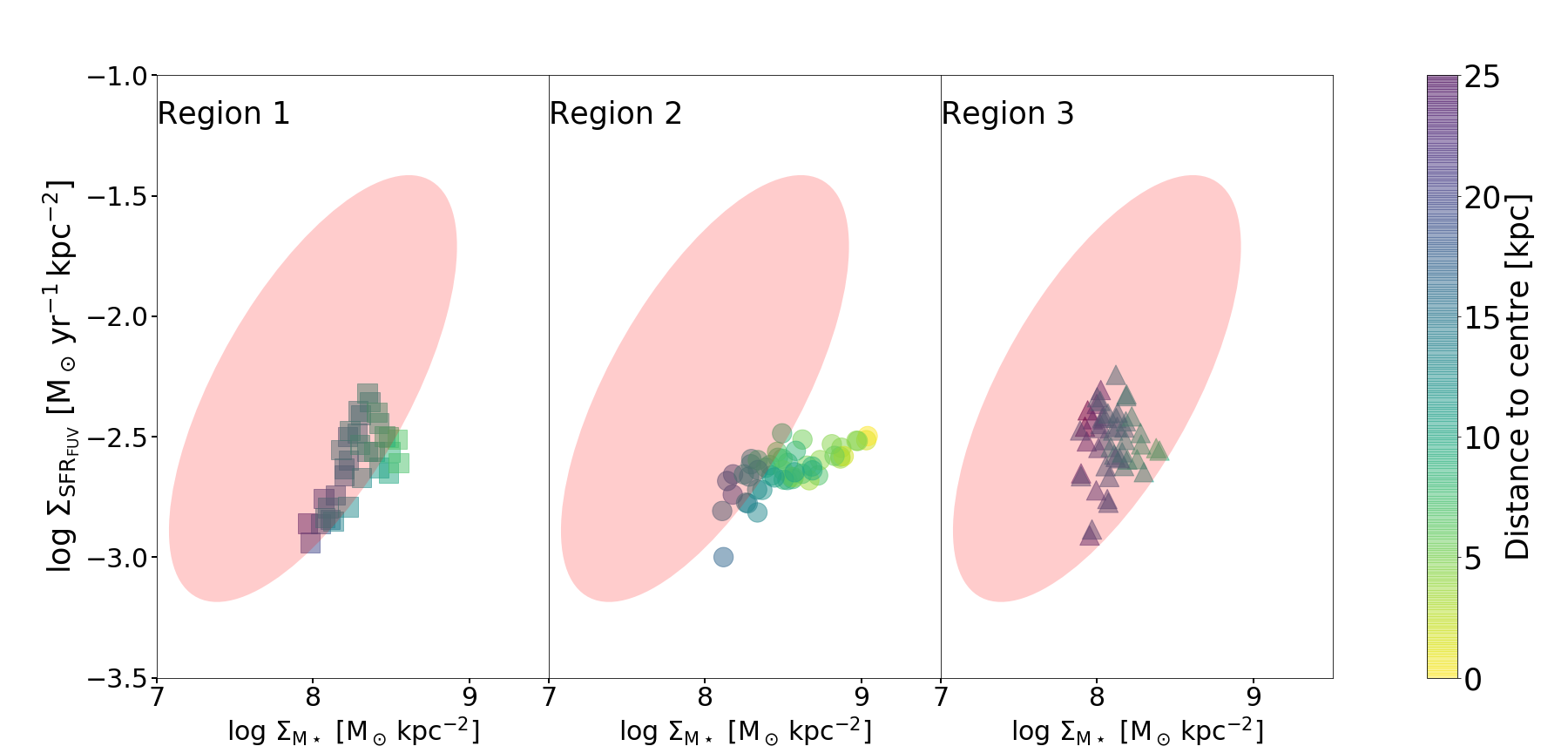}
    \centering
    \includegraphics[width=\textwidth]{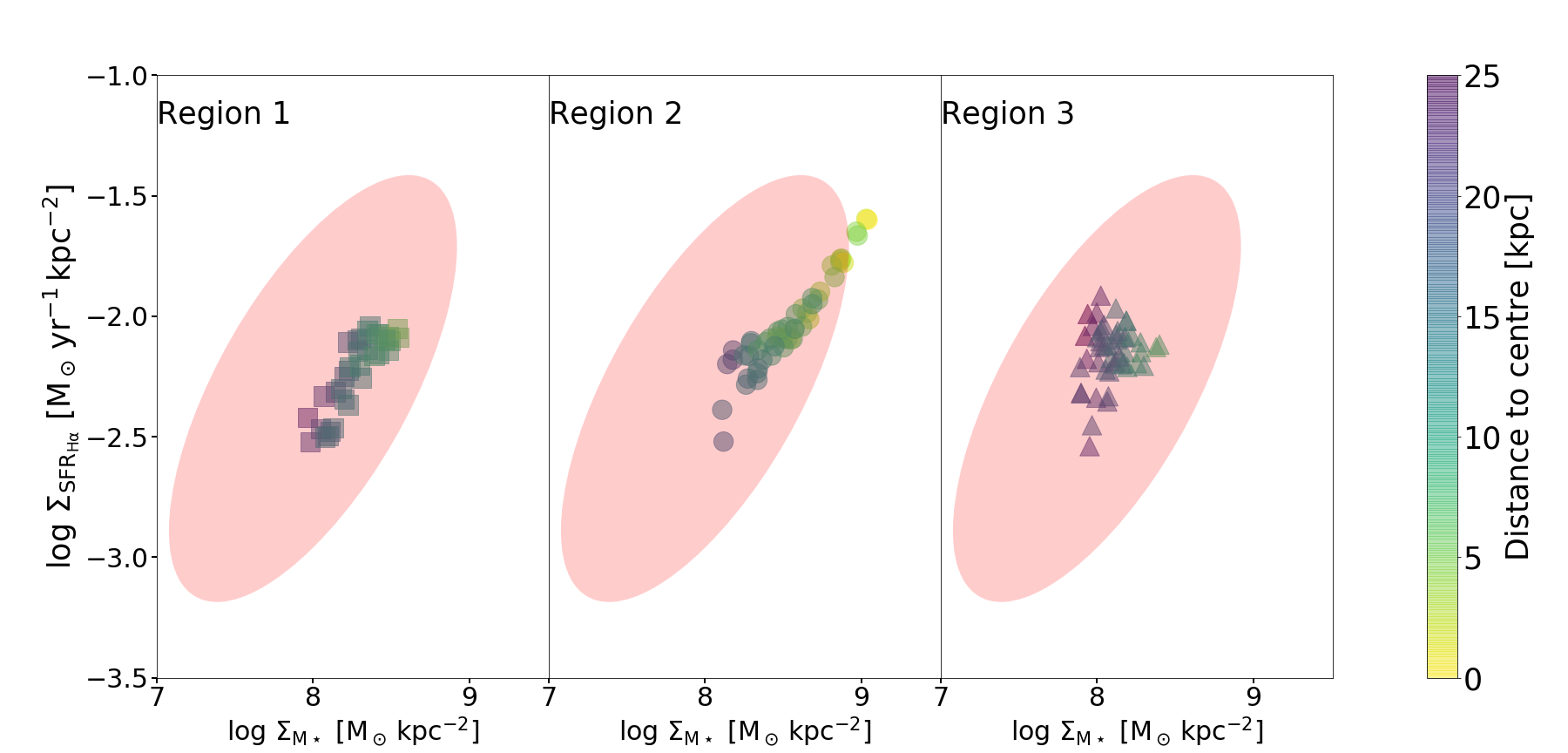}
  \caption{SFR per area versus stellar mass surface density  for different regions in the disk of 
  UGC~10214. The upper row shows the SFR derived from the FUV and 24~\mi\ emissions, \sfrfuv, and the lower row the SFR derived from the \halpha\ and 24~\mi\ emission, \sfrhalpha. The pixels in the separate panels correspond to the different regions shown in Fig.~\ref{fig:insets}. The colour of the markers of the different regions indicates the distance to the galaxy centre. For comparison the pink ellipse shows the location of the resolved MS from \citet{ellison18}.}
   \label{fig:sfr_stars}
\end{figure*}

Figure~\ref{fig:sfr_stars} shows the surface density of \sfrfuv, and \sfrhalpha , as a function of stellar mass surface density for the three regions, colour-coded with the radial distance from the centre. Overlaid as a pink ellipsoid is the spatially resolved (on kpc scale) MS found by \citet{ellison18} for a sample of 1390 star-forming galaxies from the Sloan Digital Sky Survey Mapping Galaxies at Apache Point Observatory (MaNGA) survey. Similar to the results from the previous section we find that (i) \sigmasfrfuv\ is systematically lower than \sigmasfrhalpha\ for a given \sigmamstar, (ii) the relation between \sigmasfr\ and \sigmamstar\ has a similar shape for the FUV and \halpha-traced SFRs in regions 1 and 3, but not in region 2 where the FUV relation is much flatter and (iii) the ratio \sigmasfrfuv/\sigmasfrhalpha\ is particularly low in the central region.

In addition to these results, we can draw further conclusions from a more detailed comparison of \sigmasfr\ with \sigmamstar\ and in particular with the MS found for SF galaxies. The \halpha-traced relation falls almost entirely within the range of the MS found by \citet{ellison18}. Only the most central regions lie outside, but these fall outside the stellar surface density range covered by the \citet{ellison18} sample so that a direct comparison is not possible. The situation is different for the FUV-traced relation. In region 2  almost all pixels, except those at the largest radial distances, have \sigmasfrfuv\  below the MS-value for the corresponding \sigmamstar. Also in region 1 the innermost pixels are slightly below the MS. In region 3, all pixels lie within the MS. Here, the sSFR (\sigmasfrfuv/\sigmamstar) is about  a factor 2 higher than in region 1.

Taken together, these results indicate that UGC~10214 is currently forming stars at a rate corresponding to MS galaxies, in practically the entire galaxy. For the recent SFR, traced by FUV, the picture is more heterogenous. Whereas the SF in region 3 is totally within the MS, and in region 1 slightly below the MS for the innermost points, it is well below the MS in region 2, indicating a supression of the SF especially in the central part.

\section{Discussion}
\subsection{Timescale of the SF tracers}
\label{sec:sf_history}

Our data set and analysis have revealed very noticeable differences between the values and  the distribution of the \halpha\ and FUV traced SFRs which are likely due to temporal changes in the SFR. The exact timescales over which each tracer probes the SFR are difficult to quantify. They depend on the definition of this timescale and on the SF history. \citet{boquien14} investigated this issue for different SF histories and found that the  \halpha\ emission, produced byionising stars, has a luminosity-weighted age of those stars that are contributing to \halpha\  of 2-4\,Myr (for a constant SFR during the past 100\,Myr or 1\,Gyr). The luminosity-weighted age of the FUV emission is longer, $\sim$15\,Myr, for a constant SFR during the past 100\,Myr, and 50\,Myr for a constant SFR during  the past Gyr. Adopting as the timescale the moment when an instantaneous burst is down to 10\% of its initial luminositynone obtains a timescale for the FUV of 100\,Myr and for the \halpha\ emission of 10\,Myr \citep[][their Tab. 2]{kennicutt12}.Thus, whereas the timescale for the \halpha\ emission is always  $\leq$ 10\,Myr, the timescale for the FUV depends on its definition and on the SF history and can take values between several tens of Myr to 100\,Myr.

We use  hybrid SF tracers by including the 24 \mi\ emission, thus the 24~\mi\ timescale has also to be taken into account. \citet{boquien14} derived luminosity averaged ages of the total infrared (TIR) luminosity of 15\,Myr (for a constant SFR over the past 100\,Myr) and of 100\,Myr (for a constant SFR over the last Gyr) which is similar to the timescale of the FUV. The 24~\mi\ emission, emitted mostly from very small dust grains heated by massive stars might have a shorter timescale. Thus, the timescale of the hybrid \halpha+24~\mi\ emission might be longer than that of \halpha\ alone. However, the 24~\mi\ emission contributes only 20\% or less (depending on the region) to \sfrhalpha, so that we do not expect it to have a large effect on the timescale. The timescales of the FUV and 24~\mi\ are (within the uncertainties) similar.

We therefore conclude that the \halpha+24~\mi\ emission traces the current SFR on a timescale of $\lesssim$ 10\,Myr, whereas FUV+24~\mi\ traces the recent SFR on a timescale of several tens of Myr.

\subsection{The effect of the minor interaction on UGC\,10214}

The interaction with the dwarf galaxy VV\,29c has had a very noticeable effect on the morphology of UGC\,10214. A long tidal tail has formed, containing a large fraction of the atomic gas ($\sim$40\%) as well as bright young stellar clusters. In addition, the morphology of the stellar disk is disturbed and the gas distribution, both atomic and molecular, is asymmetric.

Our data set and analysis have revealed very noticeable changes of the distribution and rate of the SF in the past tens of Myr. After a period of suppressed SF in the central region several tens of Myr ago, the SFR has increased globally and changed its location from the outer spiral arms where bright knots are visible in the FUV and 24~\mi\ emission to the centre where the most active SF is currently taking place. 

Have these changes of the SF rate and distribution been caused by the minor interaction? Following \citet{tran03}, we can estimate when the tail forming event took place, by dividing the length of the tidal tail ($\sim$100\,kpc) by the expected gas ejection velocity, which we adopt as similar to the galaxy rotation velocity (about 400\,\kms\ from the HI), yielding $\sim$250\,Myr. Thus, from a timescale point of view, the minor interaction might very well have caused the changes in the SF activity. The presence of star clusters, some of them being very bright (see the supercluster found and analysed by \citealt{tran03}) in the disk and tail, that are younger than this dynamical timescale also suggests that their formation has  possibly been triggered by the event. The minor merger is very likely responsible for the unusual distribution of the SF in the disk, with bright SF knots, visible in 8~\mi, 24~\mi\ , and FUV  in the spiral arms. The ring-like structure in the western arm might indicate, as suggested by \citet{jarrett06}, that the interaction was an off-centre collisional interaction. The current (moderate) increase in the SF in the centre, visible in \halpha, might have been caused by a gas flow towards the centre. The thin CO filament connecting the eastern side to the nucleus, visible in the CARMA map, is a hint for this, together with the broad IRAM CO spectra (widths of 500\,\kms) at the corresponding positions 2 and 3.
 
The gas properties, on the other hand, do not seem to be considerably affected by the past interaction. The pv diagram show a regular rotation curve with no indication of a major perturbation, even though at the higher velocity resolution of the IRAM spectra, the line shapes are unusual at some positions. The fact that the average \halpha-based molecular gas depletion time is  normal  fits into this picture and shows a galaxy in which the star formation seems to proceed in the same way as  in undisturbed spiral galaxies.

\subsection{Evolutionary phase of UGC\,10214}

\begin{table}
\caption{Optical, IR colours and mass ratios of UGC\,10214}           
\label{tab:colours_mass_ratios}      
\centering          
\begin{tabular}{ll}
 \hline 
 $u'_{\rm corr}$ \tablefootmark{a}&  2.73\,mJy\\
$r'_{\rm corr}$\tablefootmark{a} &  18.43\,mJy\\
$(u'-r')_{\rm corr}$ & 2.1~mag \\

\mmol/\mhi &  $0.40 \pm 0.07$\\
\mmol/\mstar & $0.07 \pm 0.01$\\
\mgas/\mstar & $0.24 \pm 0.02$  ($0.34 \pm 0.02$)\tablefootmark{b}\\
\hline
\end{tabular}
\tablefoot{
\tablefoottext{a}  {Corrected for Galactic extinction and internal extinction from CIGALE. }
\tablefoottext{b}  {Including the atomic masses of the disk and the tails ($1.2\,\times 10^{10}$ \msun) (Knierman et al. in prep.)}
}
\end{table}

UGC\,10214 is a massive galaxy, close to the upper end of the mass distribution of galaxies. Our multi-wavelength data set allows us to determine its evolutionary stage it is and make predictions about its future evolution.

In order to determine the location of UGC\,10214 on the galaxy MS, we compare its SFR and stellar mass to  a sample consisting of 12\,006 SDSS galaxies with \mstar\ $> 10^{10}$\,\msun\ used in \citet{saintonge16} as the parent distribution for their COLDGASS study. This mass range is adaquate for our object which is important because the relation between SFR and \mstar\ flattens considerably above galaxy masses of about 10$^{10}$\,\msun, \citep{noeske07, schreiber16}, i.e  the SFR per stellar mass is decreasing for high stellar masses. The reason for this flattening is not clear. We find from a comparison to Fig. 1 and Eq.~5 of \citet{saintonge16} that UGC\,10214 falls  into the range of MS SF galaxies, with \sfrfuv\ lying right on the MS and \sfrhalpha\ a factor 2.5 above. Thus, in spite of the minor merger, UGC\,10214 has a SFR that is normal for its mass.

The extinction-corrected optical colour $u^\prime-r^\prime$ (see Table.~\ref{tab:colours_mass_ratios}), corrected for internal extinction, locates UGC\,10214 in the middle of the green valley \citep{alatalo14, schiminovich07}. This apparent contradiction (lying on the star-forming MS but at the same time in the optical green valley) has its explanation most likely in the high stellar mass of UGC\,10214 and is not unusual. \citet{cortese12} found that red spirals, selected based on their optical colours were not quiescent but formed stars at a normal ($>1$ \msun\,yr$^{-1}$) rate and showed UV-optical colours typical for the blue cloud. They concluded that at high stellar masses, optical colours do not allow to distinguish between actively SF and quiescent objects. This is also the case for UGC\,10214 which has relatively red optical colours due to the bulk of stars formed in earlier times which dominate the optical colors independent of the present SFR. The low sSFR (see Tab.~\ref{tab:results}) reflects this as well. Additionally, the minor merger which has most likely produced an increase in the SFR in the past might have moved UGC~10214 from below right on the starforming MS.

Thus, UGC\,10214 is a massive spiral galaxy with a normal SFR and molecular depletion time. As discussed in the previous section, the minor merger has most likely caused a moderate increase of the SFR but has not perturbed the gas significantly. The total gas fraction (\mgas/\mstar) and molecular gas fraction (\mmol/\mhi) are normal for a galaxy of its mass \citep[see][]{saintonge16}. The gas reservoir is enough to continue SF for another 5 -- 10\,Gyr with increasing  stellar mass significantly in the future. 

\section{Summary and conclusions}

We present a multiwavelength analysis of the disk galaxy UGC\,10214 that has experienced a minor interaction with the dwarf galaxy  VV~29c about 250\,Myr ago. These data allow us to measure the total atomic gas mass, dust mass, molecular gas mass and surface density, the SFRs on different time-scales (based on the \halpha\ and FUV emissions) and the stellar mass. In addition, we carry out SED fittings with CIGALE for the entire disk and for three different regions. Our results can be summarised as the following:

\begin{itemize}

\item UGC~10214 is a massive galaxy, \mstar = $\left(1.28\pm0.13\right)\times10^{11}$~\msun, with a moderate current SFR, \sfrhalpha = $\left(4.67\pm0.19\right)$~\msun\ yr$^{-1}$.

\item We measure the total atomic, $\left(2.21\pm0.07\right)\times10^{10}$~\msun, and molecular, $\left(9.0\pm1.5\right)\times10^{9}$~\msun, gas mass of UGC~10214. The molecular gas fraction,  \mmol/\mhi\ = $0.40\pm 0.07$, shows that UGC~10214 is a molecular gas rich galaxy. The total gas fraction, \mgas/\mstar\ = $0.2\pm0.03$, and the gas-to-dust mass ratio is 180 (including He), similar to the Milky Way value.

\item We measure the atomic gas mass of the companion dwarf galaxy VV\,29c, $\left(3.5\pm 0.2\right)\times10^9$\,\msun. We do not detect any CO emission at the position and velocity of VV\,29c with an upper limit of the molecular gas mass as $10^8$\,\msun. The molecular gas fraction in this object is thus low ($< 0.03$) which is  typical for a dwarf galaxy.

\item The molecular and atomic gas have an irregular distribution over the disk of UGC~10214. The general gas kinematics, on the other hand, shows a regular rotation pattern with no visible indications of major perturbations.

\item The comparison of the SFR derived from the FUV, \sfrfuv, and from the \halpha\ emission, \sfrhalpha, reveals that the distribution of the SF has changed drastically in the recent past. Whereas  \sfrfuv\ is concentrated in the outer spirals arms, \sfrhalpha\  shows a pronounced peak in the centre. This shows that there has been a shift of the SFR from the outer disk to the centre in the past  10--100\,Myr.

\item The SFR has increased globally by a factor of 2--3  in the recent past. This increase has been particularly strong in the centre (factor of $\sim$4) where a peak in the current star formation is found. Whereas the current  SFR lies everywhere within the spatially resolved MS, the recent, FUV-derived SFR, in the central regions is below the MS.

\item The current molecular gas depletion time, \taudephalpha, averaged over the entire disk, is $\left(1.9\pm0.4\right)$\,Gyr, similar to the mean value found for spiral galaxies.

\item The optical show that UGC~10214 is in the optical green valley but its SFR follows the galaxy SF main sequence.
The most likely reason for this apparent discrepancy is the high stellar mass which dominates the optical colors.

\end{itemize}

These results show that UGC\,10214 is a massive spiral galaxy with a normal (with respect to its stellar mass) SFR and a normal molecular gas depletion time. The minor merger has most likely caused variations in the SFR during the past $\sim$Myr and a moderate increase in the SFR. It does not seem to have  perturbed the gas significantly at the present time.

\begin{acknowledgements}
DRB acknowledges financial support from the Spanish Ministry of Economy and Competitiveness (MINECO) under the grant number AYA2016-76219-P.
UL and SV acknowledge support by the research projects
AYA2014-53506-P and AYA2017-84897-P from the Spanish Ministerio de Econom\'\i a y Competitividad,
from the European Regional Development Funds (FEDER)
and the Junta de Andaluc\'ia (Spain) grants FQM108.

KK is supported by an NSF Astronomy and Astrophysics Postdoctoral Fellowship under award AST-1501294.

MB acknowledges support from FONDECYT regular grant 1170618.

We acknowledge the usage of the HyperLeda database (http://leda.univ-lyon1.fr) and of the
Nasa Extragalactic Database (NED, https://ned.ipac.caltech.edu).

Based on observations made with the NASA \textit{Galaxy Evolution Explorer}. \textit{GALEX} is operated for NASA by the California Institute of Technology under NASA contract NAS5-98034.

Funding for SDSS-III has been provided by the Alfred P. Sloan Foundation, the Participating Institutions, the National Science Foundation, and the U.S. Department of Energy Office of Science. The SDSS-III web site is http://www.sdss3.org/. SDSS-III is managed by the Astrophysical Research Consortium for the Participating Institutions of the SDSS-III Collaboration including the University of Arizona, the Brazilian Participation Group, Brookhaven National Laboratory, University of Cambridge, Carnegie Mellon University, University of Florida, the French Participation Group, the German Participation Group, Harvard University, the Instituto de Astrofisica de Canarias, the Michigan State/Notre Dame/JINA Participation Group, Johns Hopkins University, Lawrence Berkeley National Laboratory, Max Planck Institute for Astrophysics, Max Planck Institute for Extraterrestrial Physics, New Mexico State University, New York University, Ohio State University, Pennsylvania State University, University of Portsmouth, Princeton University, the Spanish Participation Group, University of Tokyo, University of Utah, Vanderbilt University, University of Virginia, University of Washington, and Yale University.

The \textit{Spitzer} Space Telescope is operated by the Jet Propulsion Laboratory, California Institute of Technology, under contract with the National Aeronautics and Space Administration.

SPIRE has been developed by a consortium of institutes led by Cardiff University (UK) and including Univ. Lethbridge (Canada); NAOC (China); CEA, LAM (France); IFSI, Univ. Padua (Italy); IAC (Spain); Stockholm Observatory (Sweden); Imperial College London, RAL, UCL-MSSL, UKATC, Univ. Sussex (UK); and Caltech, JPL, NHSC, Univ. colourado (USA). This development has been supported by national funding agencies: CSA (Canada); NAOC (China); CEA, CNES, CNRS (France); ASI (Italy); MCINN (Spain); SNSB (Sweden); STFC, UKSA (UK); and NASA (USA).

This work is based on observations carried out under project number 062-05 with the IRAM 30m telescope. IRAM is supported by INSU/CNRS (France), MPG (Germany) and IGN (Spain).

The National Radio Astronomy Observatory is a facility of the National Science Foundation operated under cooperative agreement by Associated Universities, Inc.

IRAF is distributed by the National Optical Astronomy Observatory, which is operated by the Association of Universities for Research in Astronomy (AURA) under cooperative agreement with the National Science Foundation \citep{1993ASPC...52..173T}

This research made use of \textsc{Astropy}, a community-developed core \textsc{Python} ({\tt http://www.python.org}) package for Astronomy \citep{2013A&A...558A..33A, 2018arXiv180102634T}; \textsc{ipython} \citep{PER-GRA:2007}; \textsc{matplotlib} \citep{Hunter:2007}; and \textsc{NumPy} \citep{:/content/aip/journal/cise/13/2/10.1109/MCSE.2011.37}.

\end{acknowledgements}

\bibliographystyle{aa}
\bibliography{biblio_tadpole}

\end{document}